\documentclass[fleqn,usenatbib]{mnras}

\usepackage{newtxtext,newtxmath}

\usepackage[T1]{fontenc}

\DeclareRobustCommand{\VAN}[3]{#2}
\let\VANthebibliography\thebibliography
\def\thebibliography{\DeclareRobustCommand{\VAN}[3]{##3}\VANthebibliography}

\usepackage{graphicx}	
\usepackage{amsmath}	

\newcommand{\lgal}{\textsc{L-Galaxies}}

\title[Clustering in MTNG]{The MillenniumTNG Project: The large-scale clustering of galaxies}

\author[S. Bose et al.]{%
Sownak Bose$^{1}$\thanks{E-mail: sownak.bose@durham.ac.uk (SB)}
Boryana Hadzhiyska,$^{2,3,4}$
Monica Barrera$^{5}$
Ana Maria Delgado$^{2}$,
Fulvio Ferlito$^{5}$,
\newauthor%
Carlos Frenk$^{1}$,
C\'esar Hern\'andez-Aguayo$^{5,6}$,
Lars Hernquist$^{2}$,
Rahul Kannan$^{2}$,
R\"udiger Pakmor$^{5}$,
\newauthor%
Volker Springel$^{5}$,
and Simon D. M. White$^{5}$
\\%
\\%
$^{1}$Institute for Computational Cosmology, Department of Physics, Durham University, South Road, Durham, DH1 3LE, UK\\%
$^{2}$Harvard-Smithsonian Center for Astrophysics, 60 Garden St, Cambridge, MA 02138, USA\\%
$^{3}$Miller Institute for Basic Research in Science, University of California, Berkeley, CA, 94720, USA\\%
$^{4}$Physics Division, Lawrence Berkeley National Laboratory, Berkeley, CA 94720\\%
$^{5}$Max-Planck-Institut f\"{u}r Astrophysik, Karl-Schwarzschild-Str. 1, 85748, Garching, Germany\\%
$^{6}$Excellence Cluster ORIGINS, Boltzmannstrasse 2, 85748 Garching, Germany
}

\date{Accepted XXX. Received YYY; in original form ZZZ}

\pubyear{2022}

\begin{document}
\label{firstpage}
\pagerange{\pageref{firstpage}--\pageref{lastpage}}
\maketitle

\begin{abstract}
Modern redshift surveys are tasked with mapping out the galaxy distribution over enormous distance scales. Existing hydrodynamical simulations, however, do not reach the volumes needed to match upcoming surveys. We present results for the clustering of galaxies using a new, large volume hydrodynamical simulation as part of the MillenniumTNG (MTNG) project. With a computational volume that is $\approx15$ times larger than the next largest such  simulation currently available, we show that MTNG is able to accurately reproduce the observed clustering of galaxies as a function of stellar mass. When separated by colour, there are some discrepancies with respect to the observed population, which can be attributed to the quenching of satellite galaxies in our model. We combine MTNG galaxies with those generated using a semi-analytic model to emulate the sample selection of luminous red galaxies (LRGs) and emission line galaxies (ELGs), and show that although the bias of these populations is approximately (but not exactly) constant on scales larger than $\approx10$~Mpc, there is significant scale-dependent bias on smaller scales. The amplitude of this effect varies between the two galaxy types, and also between the semi-analytic model and MTNG. We show that this is related to the distribution of haloes hosting LRGs and ELGs. Using mock SDSS-like catalogues generated on MTNG lightcones, we demonstrate the existence of prominent baryonic acoustic features in the large-scale galaxy clustering. We also demonstrate the presence of realistic redshift space distortions in our mocks, finding excellent agreement with the multipoles of the redshift-space clustering measured in SDSS data.
\end{abstract}




\section{Introduction}

The $\Lambda$ Cold Dark Matter ($\Lambda$CDM) model is, at present, our best model for the formation of the large-scale structure of the Universe. A major reason for its popularity is its predictive power: most notably, the distribution of galaxies throughout the cosmos, the morphology of which reveals a rich network of structures in the form of filaments, nodes, walls, and voids -- the so-called ``cosmic web'' \citep{Geller1989}. The development of this model has been complemented by an equally active programme of galaxy redshift surveys; the impressive agreement -- both qualitative and quantitative -- between the results of these surveys and the $\Lambda$CDM model has helped elevate it to its status as the concordance model for cosmology \cite[e.g.][]{Colless2001,Tegmark2004,Zehavi2005,Eisenstein2005,Cole2005}.

With this being said, it is important to clarify that the $\Lambda$CDM model does not predict the clustering of the galaxy field directly. Instead, it provides a framework for predicting the density field of the dark matter following epochs of gravitational instability, settling eventually into dark matter ``haloes'' that ultimately act as the sites of galaxy formation. As haloes form preferentially in locations where the initial density fluctuations were large, they are considered to be biased tracers of the underlying density field \citep[][]{Mo1999,Sheth1999,Gao2005,Gao2007,Wechsler2006}. Furthermore, galaxies are thought to be biased tracers of dark matter haloes, adding yet another layer of complexity on top \citep[see][for a review of ``galaxy bias'']{Desjacques2018}. If we are to ultimately use the observed galaxy field to constrain our cosmological model with high precision, it necessitates the development of a robust and consistent framework for establishing the mapping between the dark and luminous matter in our Universe.

There are a number of different approaches to modelling the galaxy-halo connection theoretically \citep[see, e.g.,][for a review]{Wechsler2018}. The simplest of these involve the statistical assignment of galaxies to haloes, the most prominent of which include abundance matching \citep[e.g.][]{Mo1999,Kravtsov2004,Tasitsiomi2004,Vale2004,Behroozi2010,Reddick2013}, empirical models \citep[e.g.][]{Conroy2009,Behroozi2013,Moster2013,Rodriguez2016,Tacchella2018,Behroozi2019} and Halo Occupation Distributions \citep[HODs,][]{Peacock2000,Benson2000,Berlind2002,Zheng2005}. Each of these methods makes an assumption regarding salient features of dark matter haloes that dictate the abundance and the properties of the galaxies that eventually reside within them. A more sophisticated approach is enabled by semi-analytic models of galaxy formation, in which an attempt is made to model the physical processes determining the formation and evolution of galaxies using sets of differential equations that can be applied to halo or subhalo merger trees either created by a stochastic model or obtained directly from collisionless simulations \citep[e.g.][]{Kauffmann1993,Somerville1999, Kauffmann1999,Cole2000,Springel2001,Croton2006,Benson2012,Henriques2015,Lagos2018}.

Undoubtedly, the most accurate route towards self-consistent modelling of galaxy formation is offered by cosmological, hydrodynamical simulations. It is only recently that our computational methods have improved to the point where hydrodynamical simulations are possible both with the high resolution needed to resolve the physical processes relevant to galaxies, and with the large volumes necessary to generate large statistical samples of simulated galaxies. In the last decade or so, several such simulations have been performed and been shown to reproduce the observable properties of galaxies with reasonable accuracy. These include projects like OWLS \citep{Schaye2010}, Illustris \citep{Vogelsberger2014,Vogelsberger2014b,Genel2014}, EAGLE \citep{Schaye2015}, MassiveBlack-II \citep{Khandai2015}, SIMBA \citep{Dave2016}, Magneticum \citep{Dolag2016}, Horizon-AGN \citep{Dubois2016}, BAHAMAS \citep{McCarthy2017}, IllustrisTNG \citep{Pillepich2018}, ASTRID \citep{Bird2022,Ni2022}, and THESAN \citep{Kannan2022a}. We refer the reader to the review by \cite{Vogelsberger2020}, which provides a comprehensive review of the latest developments in this subject. Despite the advancement of this field, however, even the largest of the simulations that has been run so far -- TNG-300 -- is only able to resolve galaxies within a periodic box of volume $\left(300\,{\rm Mpc}\right)^3$ -- a far cry from the kinds of survey volumes that will be mapped with ongoing and future galaxy redshift surveys.

It is the advent of these new redshift surveys that motivates the present endeavour. Ongoing surveys like DESI \citep{Levi2013} and future missions like {\it Euclid} \citep{Laureijs2011} and the {\it Roman Space Telescope} \citep{Spergel2015} will eventually the map the large-scale structure out to several tens of Gpc$^3$. A primary target for these missions is to produce the most precise measurement of the baryonic acoustic oscillation (BAO) scale, a standard distance ruler that provides one of the cleanest measurements of the expansion rate of the Universe. This feature, imprinted in the galaxy distribution on a scale of $\sim150$ Mpc is already at the edge of what has thus far been achievable in existing cosmological hydrodynamical simulations. Realising the full potential of these surveys therefore demands new generations of simulations that are able to keep pace with the ever-increasing volume of galaxy redshift surveys. 

In this paper, we present predictions for the clustering properties of galaxies in a new set of large volume, cosmological hydrodynamical simulations, named the MillenniumTNG (or MTNG) project. As suggested by the name, the flagship simulation of this suite is one that integrates the IllustrisTNG model of galaxy formation in the 740~Mpc periodic box of the iconic Millennium simulation \citep{Springel2005}, updated to use the latest constraints on the cosmological parameters. The result is a galaxy population evolved within a volume that is nearly 15 times larger than TNG-300, with only a modest compromise in resolution. In addition to this run, we have also performed a series of dark matter-only simulations within the same volume (including a pair of runs designed to suppress cosmic variance), simulations that assess the impact of massive neutrinos with varying mass, as well as a substantially larger dark matter-only run, also including neutrinos, comprising more than 1 trillion particles in a box of size 3~Gpc. The dark matter-only members of the MTNG suite have been designed to be augmented with semi-analytic galaxies,  allowing the generation of mock galaxy catalogues, both in snapshots and on the past lightcone, distributed over length scales relevant to upcoming surveys. 

The layout of this paper is as follows. We introduce the MillenniumTNG simulation suite, describing the full physics and dark matter-only runs in Section~\ref{sec:simulations}. Our main results are presented in Section~\ref{sec:results} where, amongst other quantities, we consider the clustering properties of full physics and semi-analytic galaxies selected by stellar mass, colour, their appearance in DESI-like sample selections, and their redshift-space clustering on lightcone mocks. Finally, Section~\ref{sec:conclusions} provides a summary and discussion of the results presented in this paper.

\section{The Millennium-TNG simulations}
\label{sec:simulations}

In this section, we briefly describe the main features of the MillenniumTNG simulation suite, focussing on the set of runs considered in this work in particular (Section~\ref{sec:intro_sims}). We then describe the \lgal{} model that has been used so far to generate semi-analytic galaxy catalogues in the subset of simulations in the series that have been run with dark matter-only (Section~\ref{sec:LGal}). The resulting galaxy catalogue is used as a complementary data set to the galaxies produced in the MTNG full physics run.

\subsection{An introduction to the simulation suite}
\label{sec:intro_sims}

The MillenniumTNG (MTNG) project comprises of a large suite of cosmological dark matter-only and hydrodynamical simulations of structure formation. The flagship simulation of the hydrodynamic series, which we hereafter refer to as MTNG740, is a cubic periodic box of size 740~Mpc on a side, and follows the co-evolution of 4320$^3$ dark matter particles (resulting in an effective particle mass of $1.65~\times~10^{8}\,{\rm M}_\odot$) and 4320$^3$ gas cells, each with an initial mass of $3~\times~10^7\,{\rm M}_\odot$. The gravitational softening lengths for the dark matter and gas components, respectively, are set to 3.7~kpc and 370~pc (as the minimum value). The physics model in MTNG740 is based on the IllustrisTNG \citep[TNG,][]{Pillepich2018b,Springel2018,Marinacci2018,Nelson2018a,Naiman2018} galaxy formation model \citep{Weinberger2017,Pillepich2018,Pillepich2019,Nelson2019a,Nelson2019b}, which has been shown to reproduce reasonably realistic galaxy populations across cosmic time. One notable difference with respect to TNG is that we do not simulate the effect of magnetic fields or  track individual metal species in MTNG740, which is a choice made to conserve memory. The former, in particular, is not expected to have any significant outcomes on the properties of galaxies \citep[see, e.g.,][]{Pakmor2017}. 

In addition to the MTNG740 full physics simulation, we also consider in this paper a series of dark matter-only volumes which we use as the basis for constructing semi-analytic galaxy catalogues using the \lgal{} model (described in Section~\ref{sec:LGal}). In particular, we consider two 740~Mpc volumes with 4320$^{3}$ dark matter particles, labelled MTNG740-4320-A/B. The `A' and `B' denominations refer to the fact that the two boxes have been generated using `fixed' and `paired' initial conditions based on the variance suppression technique of \cite{Angulo2016}. In short, the two realisations, `A' and `B', have been generated by first `fixing' the initial Fourier mode amplitudes to the ensemble average power spectrum, and `pairing' them such that the initial modes are exactly out of phase. The net effect of this is to reduce the variance arising from the sparse sampling of wavemodes in any individual cosmological simulation, and to boost the statistical precision of the combined set by factors of $\sim 30-40$ \citep[e.g.,][]{Chuang2019,Ding2022}. Initial conditions for the MTNG740 full physics simulation correspond to the `A' series. We also consider a second set of paired simulations, labelled MTNG740-2160-A/B, which has 8 times poorer mass resolution than MTNG740-4320-A/B. This set is used to study the clustering of semi-analytic galaxies on the lightcone, which will be our focus in Section~\ref{sec:lightcone}. 

In all the simulations used in this work, we adopt cosmological parameters given by \cite{Planck2016}: $\Omega_0 = 0.3089$ (total matter density), $\Omega_{\rm b} = 0.0486$ (baryon density), $\Omega_\Lambda = 0.6911$ (dark energy density), $H_0 = 67.74$ kms$^{-1}$Mpc$^{-1}$ (Hubble parameter) and $\sigma_8 = 0.8159$ (linear rms density fluctuation in a sphere of radius 8 $h^{-1}$ Mpc at $z=0$). These are identical to those adopted in TNG in order to aid comparison. Initial conditions are generated at $z~=~63$ using second-order Lagrangian perturbation theory based on a modern version of the {\sc NgenIC} code \citep{Springel2005}. 

The dark matter-only simulations in this project have been run using a modified version of the publicly-available {\sc Gadget-4} code \citep{Springel2021}, while all full physics simulations have been performed using the {\sc Arepo} code \citep{Springel2010,Weinberger2020}. In order to streamline data processing, we have opted to perform common tasks like halo and substructure finding, merger tree construction, lightcone creation, and the computation of matter power spectra on-the-fly. In particular, halo finding starts by defining groups using a friends-of-friends algorithm, while hierarchical bound structures within these groups are identified using the {\sc subfind-hbt} algorithm \citep{Springel2021}. In what follows, `galaxies' are defined as collections of stars that exist within bound structures identified by {\sc subfind-hbt}. We have generated a series of halo-, galaxy-, and particle-based lightcones in a variety of geometries; for further details, we refer the reader to the  paper by \cite{Cesar2022}.

The present paper represents one of an initial series of publications associated with the MTNG project. Companion papers in this series include that by \cite{Cesar2022}, which provides a detailed introduction to the project as a whole, including all data products such as group catalogues, merger trees, and lightcones, and an analysis of the dark matter and halo clustering statistics. \cite{Pakmor2022} introduces the full physics MTNG simulations, with a particular focus on the scaling relations of rich galaxy clusters. The latest version of the \lgal{} semi-analytic model, and its application to the lightcone outputs in MTNG are presented in \cite{Barrera2022}. In \cite{Boryana2022a,Boryana2022b}, we test and improve upon standard HOD prescriptions of the galaxy-halo connection based on measurements of the galaxy population in MTNG, while in \citet{Contreras2022} we introduce a powerful inference methodology that is capable of constraining the cosmological parameters of MTNG from galaxy clustering. \cite{Delgado2022} presents a comprehensive study of the intrinsic alignments of elliptical and spiral galaxies, and the correlation of galaxy shapes with their large-scale environment. \cite{Furlito2022} examines the weak gravitational lensing signal predicted in our dark matter-only and full physics runs. Finally, in \cite{Kannan2022}, we investigate the properties of very high redshift galaxies ($z\geq8$) and consider the predictions of our model against recent observations made by the {\it James Webb Space Telescope}.

\subsection{The \lgal{} semi-analytic model}
\label{sec:LGal}

The \lgal{} semi-analytic model \citep{Springel2001,Springel2005,Croton2006,Guo2011,Yates2013,Henriques2015,Henriques2020,Ayromlou2021} is used to generate synthetic galaxy populations within dark matter-only simulations in post-processing. This is achieved through a set of coupled differential equations used to capture the galaxy formation process, and is applied to dark matter halo merger trees constructed from MTNG740-2160/4320-A/B simulations. The \lgal{} model includes prescriptions that account for gas cooling and star formation in haloes, as well as a variety of feedback modes including those from supernovae and black holes. The model is able to also account for ejection and recycling of gas from haloes, as well as the formation and evolution of metals, broad-band line luminosities etc. Galaxy catalogues generated with \lgal{}, therefore, act as a perfect foil for comparisons to the full physics MTNG simulations, particularly those that have been run with comparable box size and mass resolution. We explore this in more detail in Sections~\ref{sec:clustering_lgal}--\ref{sec:lightcone}.

The particular version of \lgal{} used in this paper is taken, together with the free parameters embedded in its physics modules, from \cite{Henriques2015}. The parameters were calibrated so as to reproduce the abundance and specific star formation rate of observed galaxies as a function of stellar mass over the redshift range $0\leq z \leq 3$. The clustering and the gas content of galaxies were not used as constraints and so provide a direct test of the model.  (For further details, we refer the reader to the companion paper by \citealt{Barrera2022}).

\section{Results}
\label{sec:results}

In this section, we will present the main results of our work. As the flagship hydrodynamics simulation in our suite, we will begin by considering the clustering properties of galaxies selected by stellar mass and colour in the MTNG740 full physics simulation (Section~\ref{sec:clustering_mtng}). Next we turn our focus to the same statistics measured in \lgal{} catalogues generated for the MTNG740-4320-A/B boxes (Section~\ref{sec:clustering_lgal}). We then consider these two sets of galaxies together and select DESI-like samples from each and consider their clustering (Section~\ref{sec:DESI}). Finally, we will conclude by considering the redshift-space clustering of \lgal{} mock catalogues generated on the MTNG lightcone outputs (Section~\ref{sec:lightcone}).

\subsection{The clustering properties of galaxies in MTNG740}
\label{sec:clustering_mtng}

\begin{figure*}
    \centering
    \includegraphics[width=\textwidth]{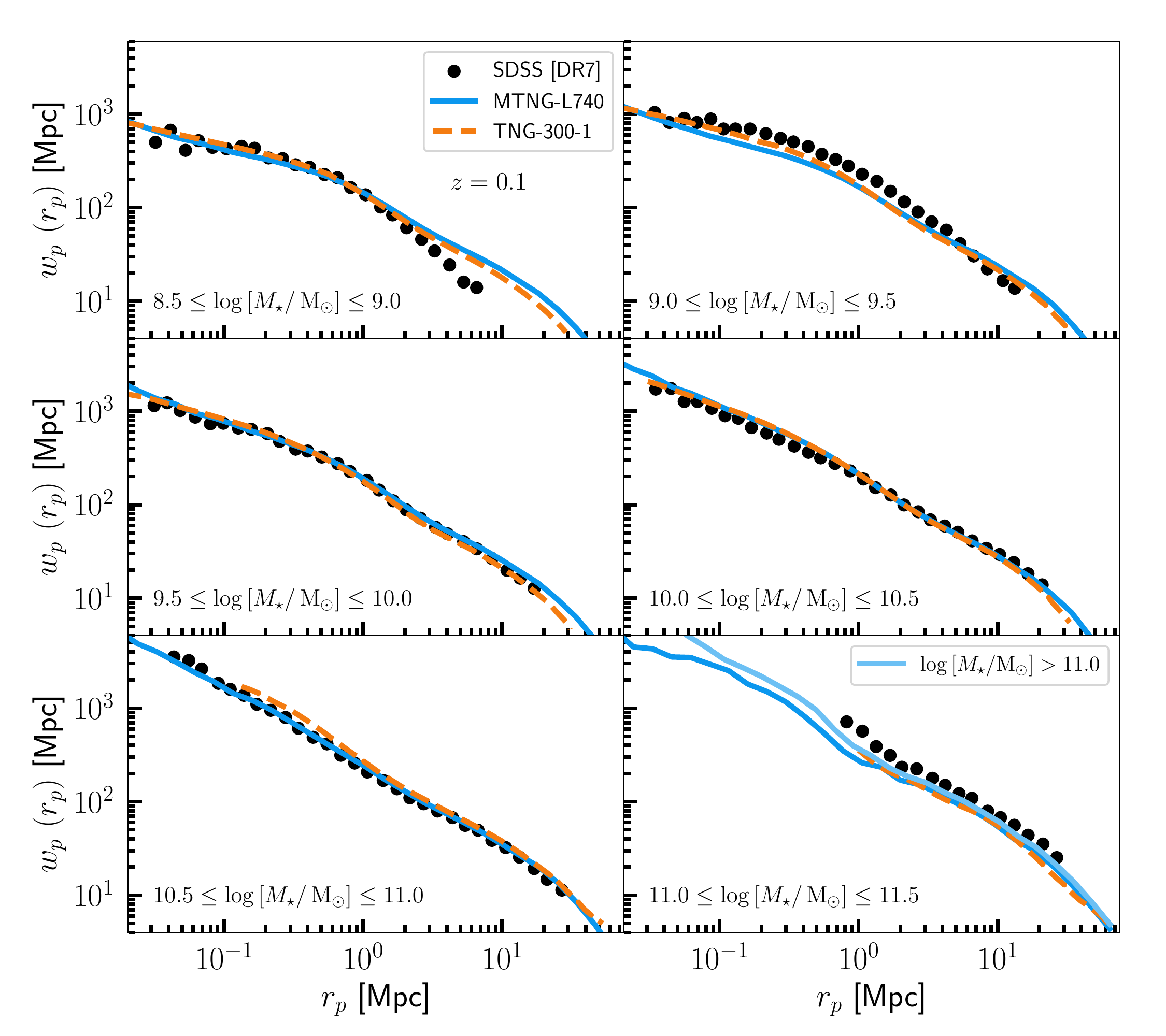}
    \caption{The projected correlation function of galaxies at $z=0.1$, $w_p$, as a function of projected separation, $r_p$. Each panel separates the galaxy population into mass bins of width 0.5 dex. The black circles show the measurements of \citet{Guo2011} from SDSS DR7 data. The solid blue line shows our measurement from MTNG740, while the orange dashed line is the same measurement made in TNG-300-1. The simulation measurements are in good agreement with each other despite the comparatively lower resolution in MTNG740 compared to TNG-300-1. The larger volume of the MTNG740 simulation allows a higher fidelity measurement of the correlation function, and this is especially evident for the most massive galaxies shown in the bottom two panels. The simulation measurements are in good agreement with the data, although there is a noticeable offset relative to the observational data points for galaxies in the range $11.0\leq\log\left[M_\star/\,{\rm M}_\odot\right]\leq11.5$ where our clustering prediction is somewhat low. The lighter blue shade in this panel shows MTNG740 predictions for $\log\left[M_\star/\,{\rm M}_\odot\right]\geq11.0$, showing the contribution of galaxies more massive than the nominal upper limit in this bin.}
    \label{fig:projected_DR7}
\end{figure*}

In order to characterise the clustering properties of galaxies in our simulations, we measure the two-point correlation function in real-space, $\xi(r)$, defined as:
\begin{equation}
    \xi(r) = \frac{DD(r)}{RR(r)} - 1,
\end{equation}
where $DD(r)$ and $RR(r)$, respectively, are the number of data-data and random-random pairs separated by some distance $r$. The full 3D distance, $r$, may be decomposed into a (transverse) projected separation, $r_p$, and a component along the line-of-sight, $\pi$, such that $r=\sqrt{r_p^2 + \pi^2}$. In observations, the more readily measured quantity is instead the {\it projected} two-point correlation function, $w_p(r_p)$, which is simply the integral of $\xi(r_p,\pi)$ along the line-of-sight:
\begin{equation}
    w_p(r_p) = 2 \int_0^{\pi_{{\rm max}}} \xi(r_p,\pi) \, {\rm  d}\pi.
\end{equation}
We set $\pi_{{\rm max}}=80\,h^{-1}$Mpc throughout and use the {\tt Corrfunc} code \citep{Sinha2017} to compute the correlation function given a set of galaxies. 

In Figure~\ref{fig:projected_DR7}, we show the projected correlation function, $w_p(r_p)$, measured from MTNG740 galaxies at $z=0.1$ selected by stellar mass. Each panel considers the clustering of galaxies with increasing stellar mass, ranging from $8.5\leq \log\left[M_\star/{\rm M}_\odot\right]\leq9.0$ in the top-left panel, to $11.0\leq \log\left[M_\star/{\rm M}_\odot\right]\leq11.5$ in the bottom-right. The results from MTNG740 are shown in blue. The black circles show the clustering measured {\it observationally} in the SDSS data as compiled by \cite{Guo2011}. Finally, the dashed orange curve shows the projected clustering measured in the TNG300-1 simulation, where the baryonic mass resolution is roughly a factor of two better than in MTNG740.

\begin{figure*}
    \centering
    \includegraphics[width=\textwidth]{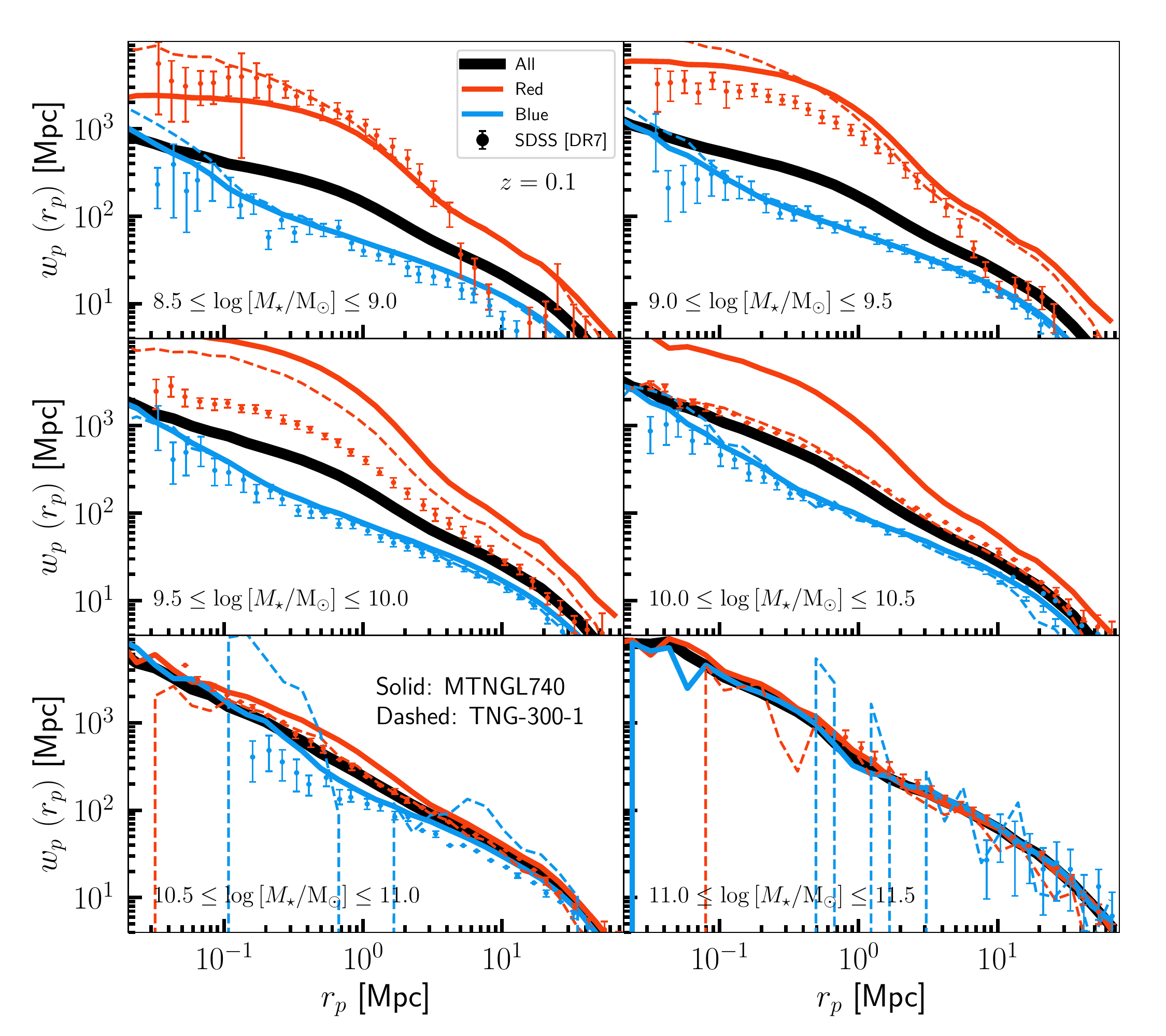}
    \caption{The projected correlation function of galaxies at $z=0.1$, $w_p$, as a function of projected separation, $r_p$. In in each panel (i.e. stellar mass bin) we have split the galaxy population into red and blue galaxies based on their $g-r$ colours (see main text for further details). The data points with errors again represent the clustering measured in SDSS DR7. The solid lines show the results from MTNG740, while the dashed lines are the measurements from TNG-300-1. While the large-scale clustering is consistent between the curves, clear differences  are apparent on scales smaller than $r_p\lesssim 1\,{\rm Mpc}$ predominantly in the clustering of red galaxies. In particular, MTNG740 predicts a excess clustering of red galaxies in the one-halo regime compared to what is measured in the data. TNG-300-1 performs slightly better, and this is especially evident in the mass range $10.0\leq\log\left[M_\star/{\rm M}_\odot\right]\leq10.5$. On the other hand, the clustering measured in MTNG740 for blue galaxies is in good agreement with the data at all masses. The measurement becomes noisy in the largest mass bin, where only a small number of galaxies are identified as blue.}
    \label{fig:projected_DR7_colour}
\end{figure*}

A first conclusion from this figure is the generally good level of agreement between the observational data and the two sets of hydrodynamical simulations. This is especially noticeable in the intermediate mass ranges ($9.5\leq \log\left[M_\star/h^{-2}{\rm M}_\odot\right]\leq11.0$). This is consistent with the conclusions of \cite{Springel2018} who quantified the clustering in IllustrisTNG (the orange curves in Figure~\ref{fig:projected_DR7} coincide with their results). While the agreement is, in general, quite good, there are regions where the observed clustering is not reproduced as well. For example, both MTNG740 and TNG-300-1 overpredict the two-halo clustering for galaxies in the range $8.5\leq \log\left[M_\star/{\rm M}_\odot\right]\leq9.0$  on scales $r_p\geq3$~Mpc. An even bigger discrepancy is observed in the largest mass bin, where now MTNG740 and TNG-300-1 underpredict the clustering strength observed in the SDSS data. Cosmic variance or finite box size effects are unlikely to be the solution here, as moving from TNG-300-1 to the $\sim15\times$ larger volume in MTNG740 makes no real difference on large scales. One source of systematic error could be in the details of how the stellar mass of the most massive galaxies are compared between observed and simulated samples. As a simple test of this, we use the light blue curve in the last panel to show the clustering of galaxies with $\log\left[M_\star/{\rm M}_\odot\right]\geq11.0$ to allow a more generous mass range for the largest galaxies in MTNG740. This slightly enhances the clustering prediction, bringing it closer to the data, but only marginally so. We leave a detailed investigation of the clustering of the most massive galaxies to future work.

It is reassuring to see that despite the lower baryonic mass resolution in MTNG740, the projected galaxy clustering is no worse than in TNG-300-1. In other words, the projected clustering by stellar mass is converged at the resolution of MTNG740. The real statistical power of MTNG740 is seen in the bottom two panels of Figure~\ref{fig:projected_DR7} which shows that we are able to make concrete predictions for the one-halo clustering of galaxies more massive than $\log\left[M_\star/{\rm M}_\odot\right]\geq10.5$, which is statistically limited in TNG-300-1. 

\begin{figure*}
    \centering
    \includegraphics[width=\textwidth]{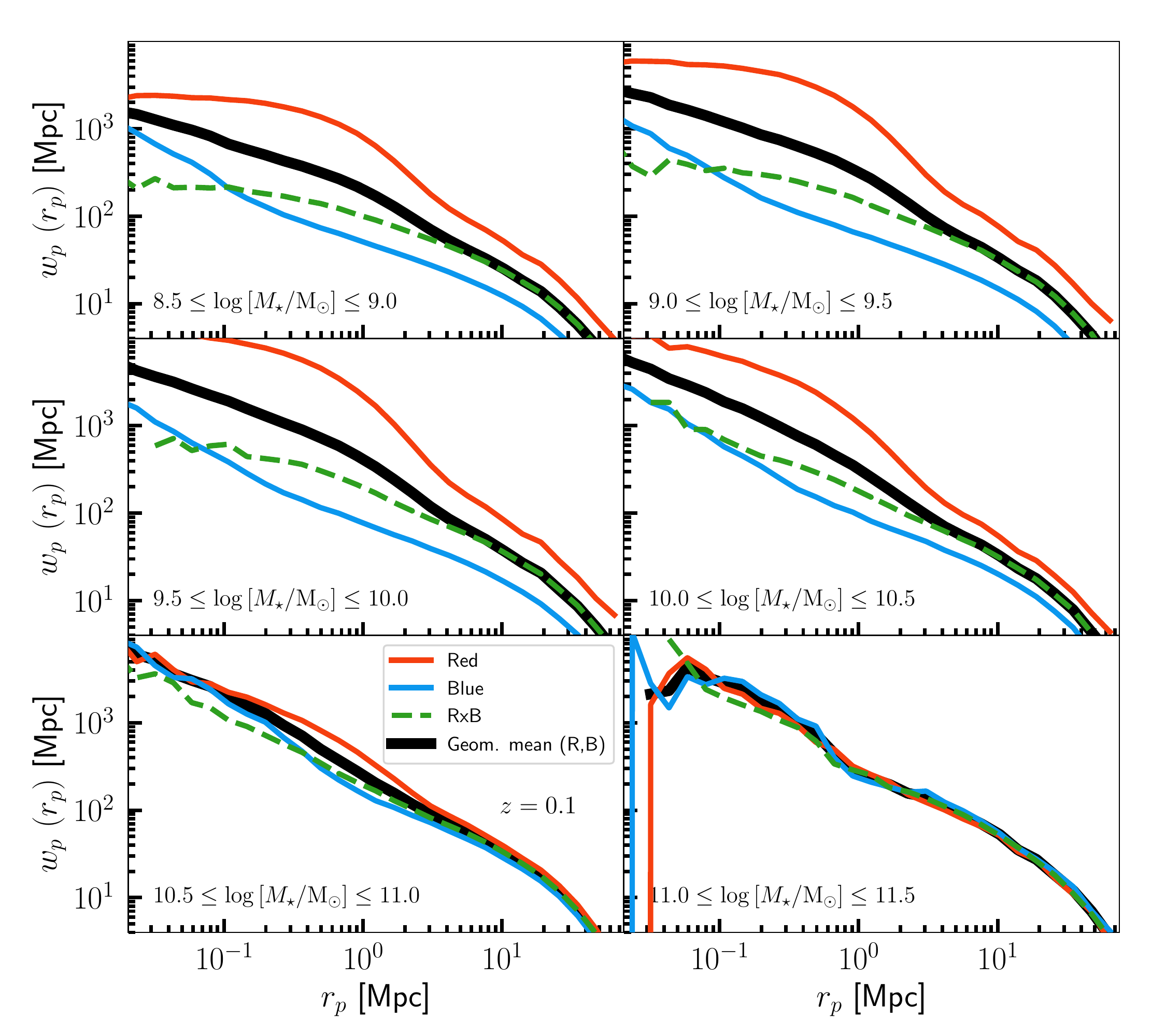}
    \caption{As Figure~\ref{fig:projected_DR7_colour}, but now also showing the {\it cross-correlation} between red and blue galaxies in MTNG740 (represented as the dashed green line). We no longer show the observational measurements for the sake of clarity. The thick black line shows the geometric mean of the red and blue galaxy auto-correlation functions. In the two-halo regime ($r_p \geq 3\,{\rm Mpc}$), the cross-correlation tracks the geometric mean which is to be expected (discussed in the main text). In the one-halo regime, the cross-correlation markedly `peels-off' from the geometric mean and is suppressed relative to it; this is due to over-abundance of red galaxies in our simulation data set.}
    \label{fig:colour_crosscorrelation}
\end{figure*}

A stronger test for any galaxy formation model is to predict the clustering of galaxies at fixed stellar mass, separated according to their star forming state. In Figure~\ref{fig:projected_DR7_colour}, we once more consider the projected clustering of galaxies selected by stellar mass in MTNG740 (again at $z=0.1$) but now galaxies in each mass bin have been segregated into `blue' and `red' galaxies based on their $g-r$ colour. To define the threshold demarcating red and blue populations, we follow \cite{Springel2018} and employ a mass-dependent colour cut:
\begin{equation}
\label{eq:cut}
    g-r = \log\left[ M_\star/h^{-1}{\rm M}_\odot \right] \times 0.054 + 0.05.
\end{equation}
Figure~\ref{fig:projected_DR7_colour} contrasts the predictions of MTNG740 (solid lines) and those from TNG-300-1 (dashed lines) with data from SDSS (points). The black curves show the clustering of all MTNG740 in each mass range. Qualitatively, our simulations are able to recover the expected trend where red galaxies are clustered more strongly than blue galaxies. In detail, however, we note interesting differences between the observed data and the predictions of the full physics simulations. While the clustering of blue galaxies is generally reproduced well in all mass bins, the clustering of red galaxies is much stronger in MTNG740 and TNG-300-1 compared to what is observed in the data. The discrepancy is especially pronounced in the intermediate mass range ($9.5\leq\log\left[M_\star/{\rm M}_\odot\right]\leq10.5$), where TNG-300-1 does a slightly better job than MTNG740. This is an arena where the better baryonic mass resolution in TNG-300-1 might result in bluer, star-forming galaxies compared to MTNG740 which displays a much higher fraction of quenched galaxies (particularly satellites). The overly strong one-halo term is an indication that too large a fraction of the red galaxies in those mass ranges are satellite galaxies in massive haloes, rather than centrals or satellites in lower mass systems. In particular, it may be the case that at the resolution of TNG-300-1 and MTNG740, ram pressure stripping may be too efficient. 

\begin{figure*}
    \centering
    \includegraphics[width=\textwidth]{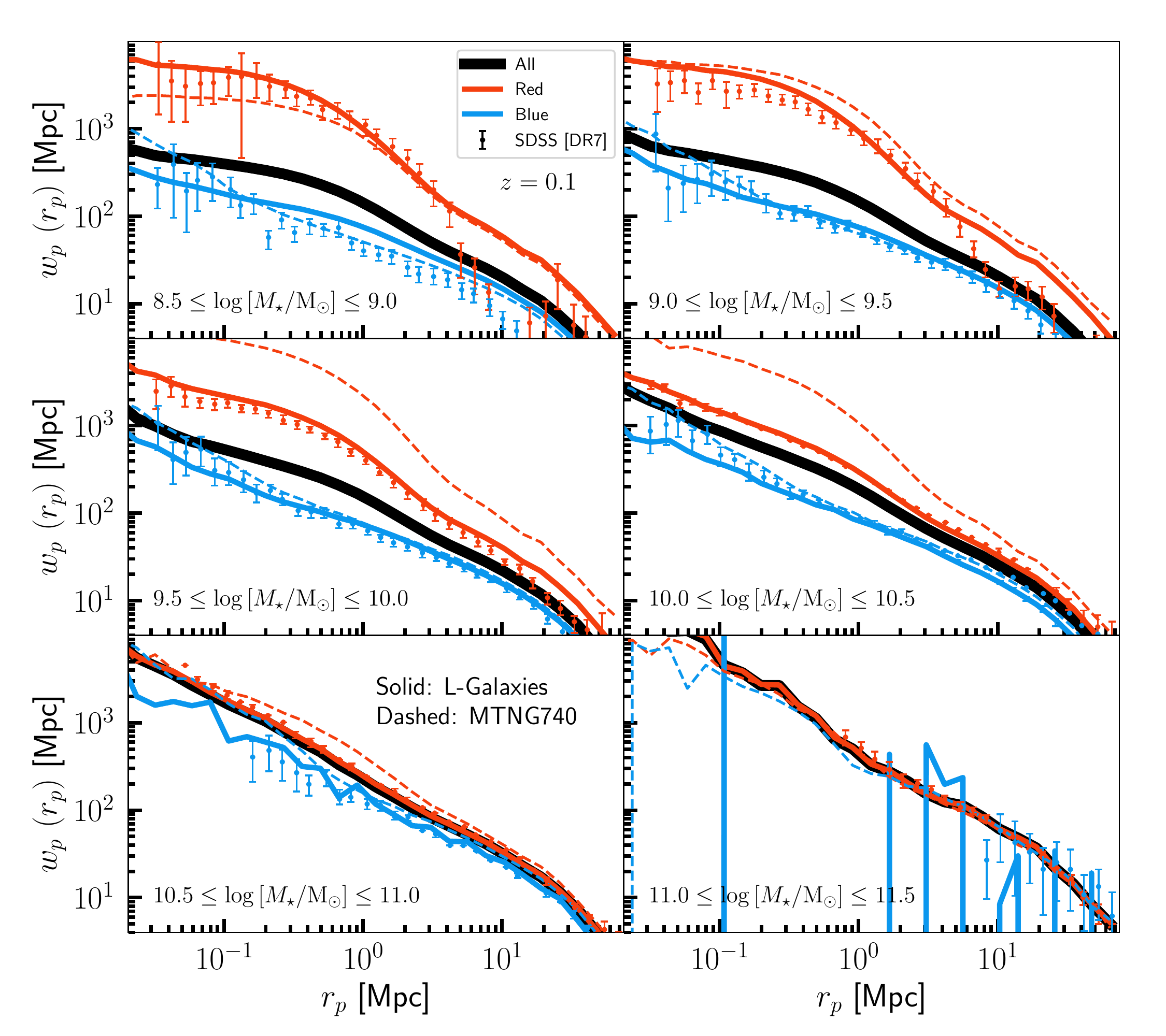}
    \caption{As Figure~\ref{fig:projected_DR7_colour}, but now showing the correlation function measured from the \lgal{} catalogue (generated from the MTNG740-A box) represented by the solid lines. For comparison, we show the results from the MTNG740 hydrodynamical simulation using the dashed curves. In general, the results from the semi-analytic galaxy catalogue agrees well with MTNG740. The main difference is seen in the middle two panels, where \lgal{} does a far better job at reproducing the clustering of red galaxies as measured in the SDSS DR7 data.}
    \label{fig:projected_DR7_colour_LGalaxies}
\end{figure*}

We examine the relative clustering of red and blue populations further in Figure~\ref{fig:colour_crosscorrelation}, which reproduces the MTNG740 curves from Figure~\ref{fig:projected_DR7_colour}, but also includes the {\it cross-correlation} of red and blue galaxies in each mass bin (shown in green). The thick black curve now shows the {\it geometric mean} of the red and blue galaxy auto-correlation functions. The cross-correlation function encodes the mixing of red and blue galaxy populations within haloes; its deviation from the geometric mean of the individual auto-correlations suggests a segregation of these populations from one another \citep[see, e.g.,][]{Zehavi2011}. Figure~\ref{fig:colour_crosscorrelation} shows that while the cross-correlation overlaps with the geometric mean on large scales ($r_p\gtrsim3$ Mpc), it is suppressed relative to the latter inside haloes. This suggests some evidence of segregation between red and blue populations within massive haloes -- which is not unexpected given that the satellites of massive haloes are expected to be predominantly red rather than blue. 

It is also interesting to note that the cross-correlation is nearly flat for $r_p\lesssim3$~Mpc in the two lowest mass bins. The absence of a contribution to the one-halo term may suggest an absence of any mixing between these populations: perhaps indicating the existence of entirely `red' or `blue' haloes. On the other hand, the cross-correlation tracks the geometric mean down to $r_p\approx300$ kpc in the two most massive bins; this is also consistent with measurements in the data \citep[see, e.g.,][]{Zehavi2005}. Slight deviations are noticeable on smaller scales, but as the number of galaxy pairs decreases on separations this small, the correlation function also becomes more noisy.

\begin{figure*}
    \centering
    \includegraphics[width=\textwidth]{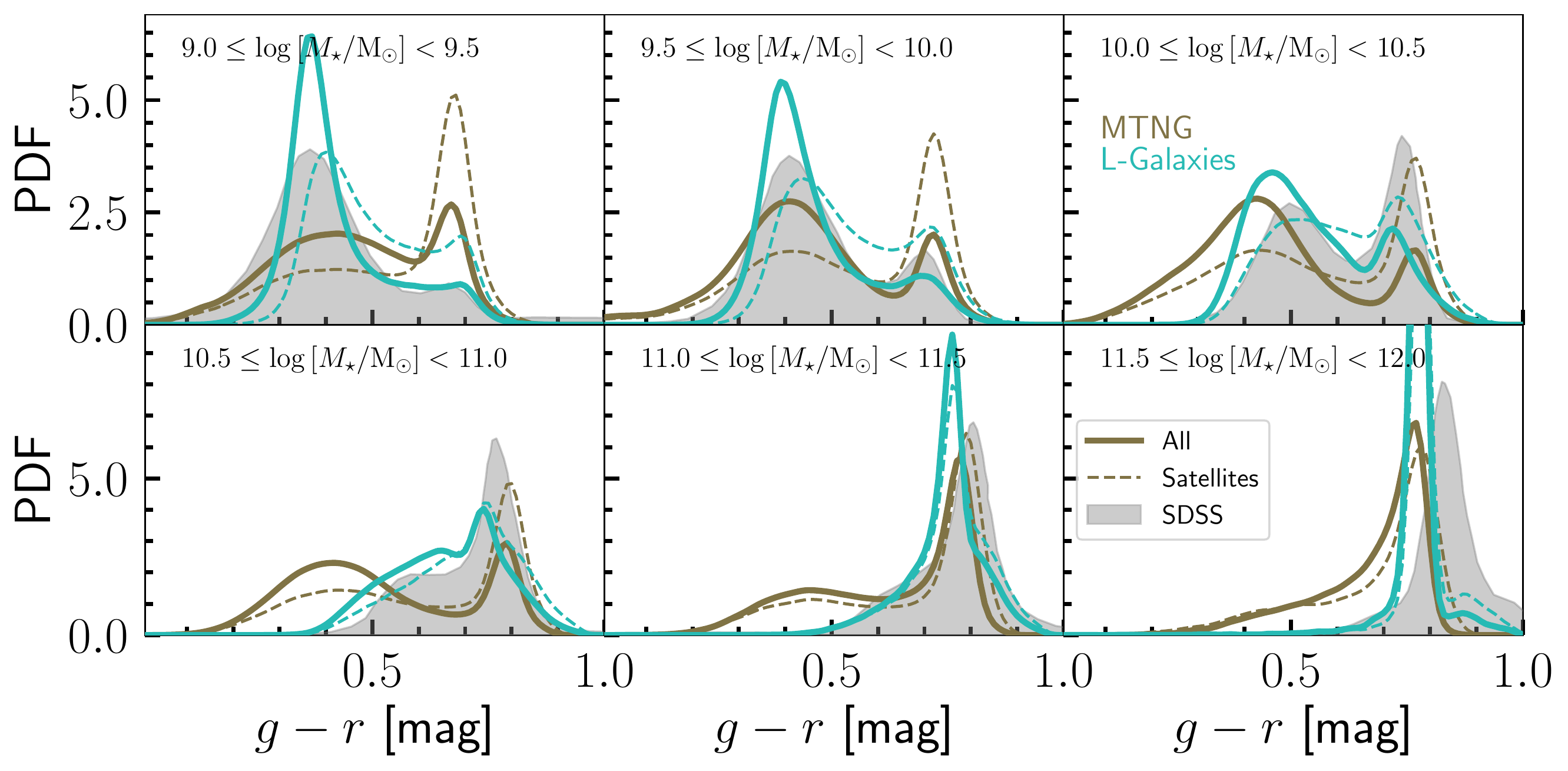}
    \caption{The distribution of $g-r$ colours as a function of stellar mass for galaxies identified at $z=0$ in MTNG740 (olive) and \lgal{} run on the MTNG740-4320-A simulation (teal). The overall galaxy populations are shown using solid lines, while the contribution of satellite galaxies alone is shown using dashed lines. Finally, the distribution of galaxy colours measured in SDSS is shown using the grey histograms. Dust extinction has only been accounted for in the \lgal{} data. There is a clear excess peak of red satellite galaxies in MTNG740 compared to \lgal{}, particularly for objects in the mass range $9.0\leq\log\left[M_\star/{\rm M}_\odot\right]\leq10.5$. Both sets of simulated galaxies otherwise show qualitative agreement with the observational data.}
    \label{fig:colour_distr}
\end{figure*}

\subsection{The clustering properties in the \lgal{} catalogue}
\label{sec:clustering_lgal}

Having established some of the basic clustering properties of galaxies in the MTNG740 full physics simulation, we now turn our attention to the performance of the \lgal{} semi-analytic model. Figure~\ref{fig:projected_DR7_colour_LGalaxies} is identical to Figure~\ref{fig:projected_DR7_colour}, except now we contrast the predictions of the MTNG740 full physics simulation (dashed lines) with the \lgal{} catalogue generated from the MTNG740-4320-A dark matter-only box (solid lines). We consider only the `A' realisation here as it encodes the same phase information as the MTNG740 full physics run, thereby easing comparison between the two sets of galaxy catalogues. We use the same threshold (Equation~\ref{eq:cut}) for determining stellar mass-dependent colour cuts in \lgal{}, where we have also corrected galaxy colours for extinction using the standard dust prescription in the model.

Figure~\ref{fig:projected_DR7_colour_LGalaxies} shows that while the projected clustering of blue galaxies is reproduced equally well in \lgal{}, the model also does a better job of predicting the correct clustering for red galaxies, particularly in the regime where MTNG740 showed large discrepancies with respect to the data. We caution here, however, that these results do not imply that the \lgal{} semi-analytic model is somehow a `better' or `more realistic' model of galaxy formation than the full physics implementation in MTNG. For instance, we know that at least part of the discrepancy in MTNG stems from numerical resolution as the more well-resolved TNG-300-1 does a better job at reproducing the colour-dependent clustering. Instead, the results show how choices made in each model result in differences in how blue and red galaxies of a given stellar mass populate dark matter haloes (i.e., resulting in different colour-dependent HODs) in \lgal{} as compared to MTNG740.

We can examine the origin for the differences in the colour-dependent clustering between MTNG740 and \lgal{} by considering the detailed distribution of galaxy colours as a function of stellar mass and satellite/central status. Figure~\ref{fig:colour_distr} shows one-dimensional kernel density weighted PDFs of galaxy $g-r$ colours at $z=0.1$ for objects identified in MTNG740 (olive curves) and \lgal{} (teal curves); these are contrasted against data from SDSS shown using the grey histograms. Satellite galaxies in MTNG740 and \lgal{} are represented by dashed curves. We find that the simulated galaxy populations as a whole (i.e. the solid curves) show reasonable qualitative agreement with the observed galaxy population, and exhibit the expected colour bimodality for galaxies with $\log\left[M_\star/{\rm M}_\odot\right]<11.0$. Above this mass we find mostly quiescent populations, signified by the predominance of the red galaxy tail. 

\begin{figure*}
    \centering
    \includegraphics[width=0.49\textwidth]{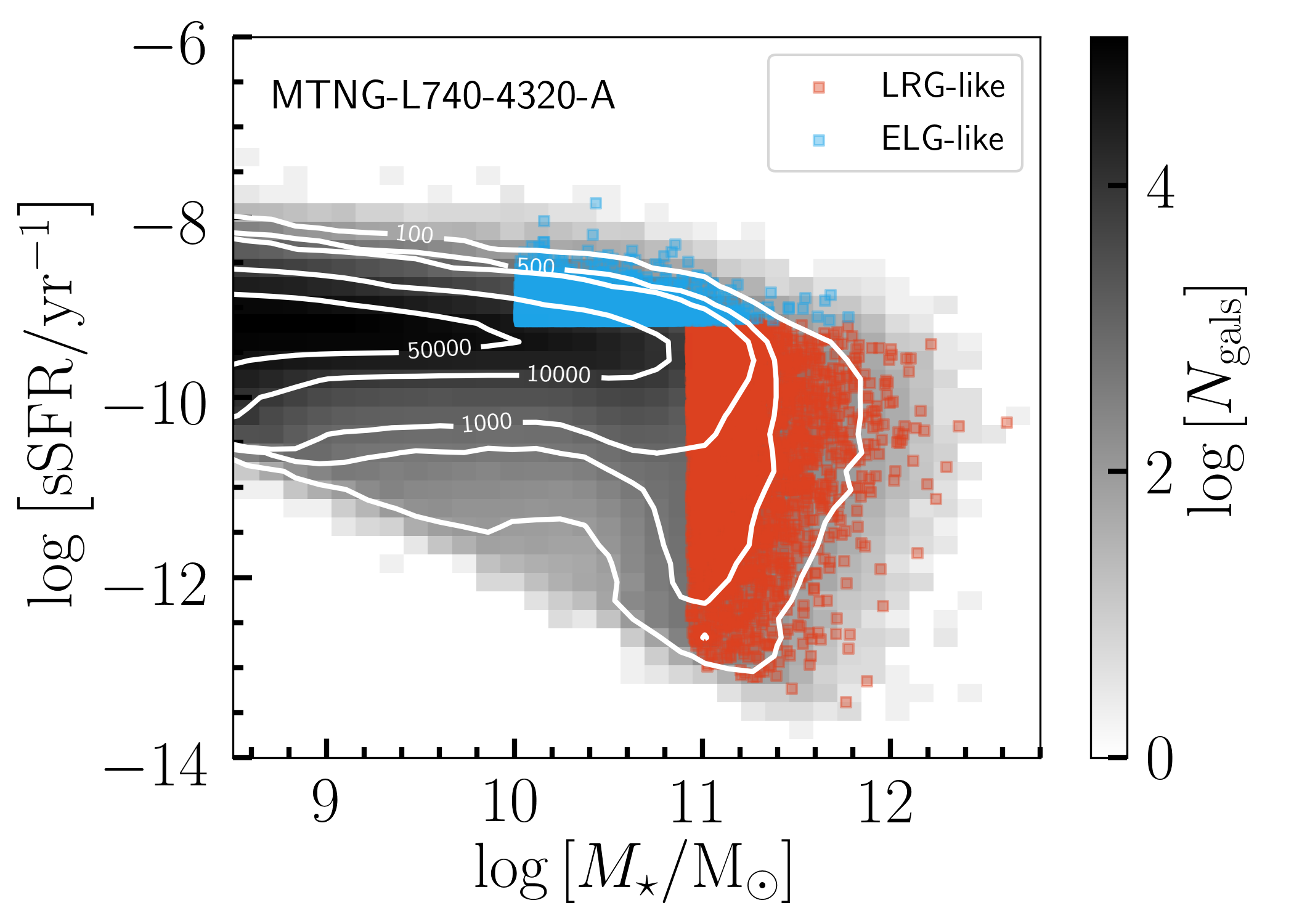}
    \includegraphics[width=0.49\textwidth]{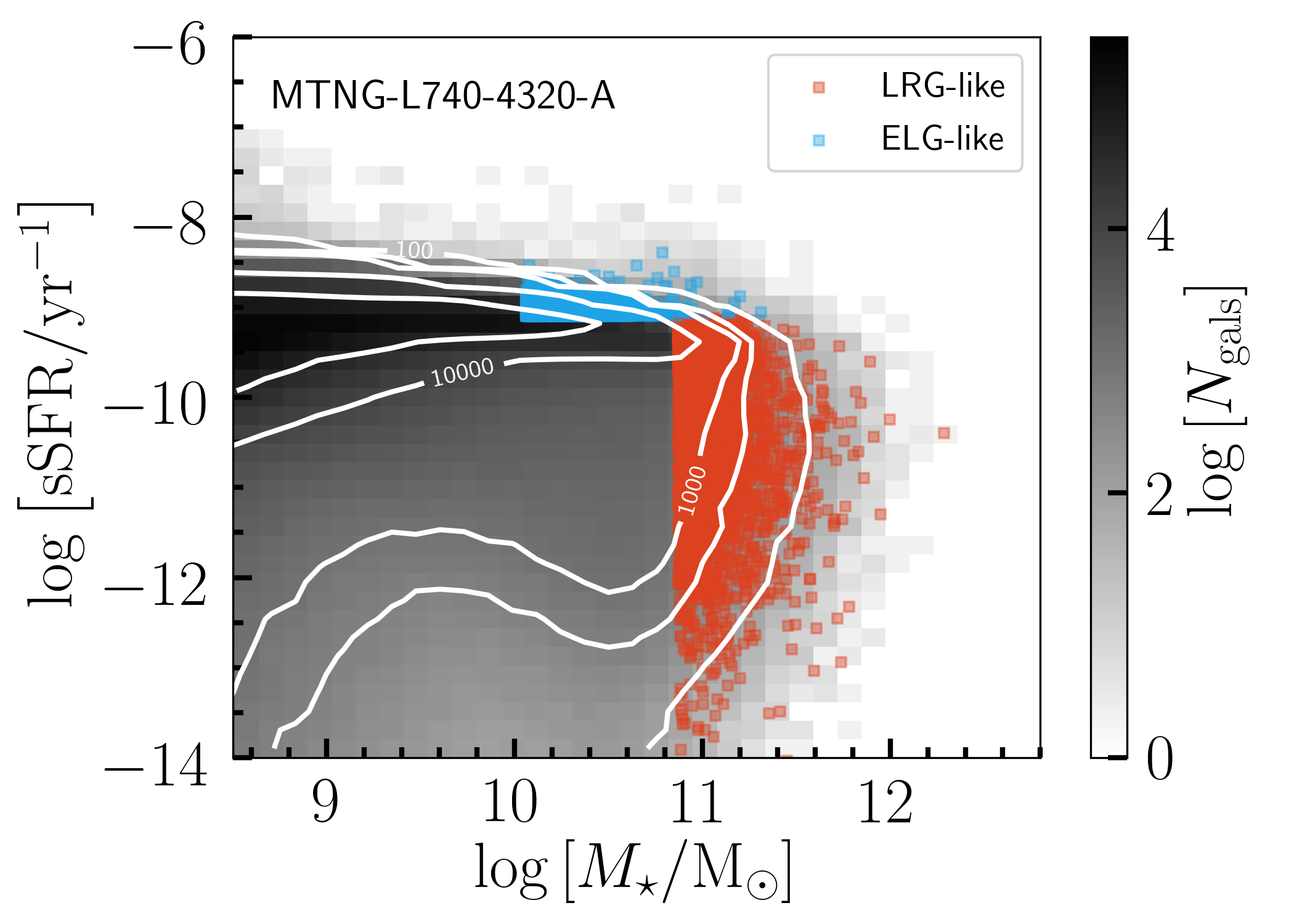}
    \caption{The distribution of galaxies at $z=0.81$ in the plane of stellar mass, $M_\star$, and specific star formation rate (sSFR) in MTNG740 ({\bf left panel}) and \lgal{} ({\bf right panel}). The greyscale histogram in the background shows the number of galaxies as denoted by the colour scale alongside each image. Red and blue points, respectively, show the subset of galaxies that pass our criteria for being identified as luminous red galaxy-like (LRG-like) and emission line galaxy-like (ELG-like) samples as might be observed by DESI. The selection criteria are described in the main text; in all cases, the galaxy samples have been generated with a target number density of $\overline{n}_{{\rm gal}} = 7\times10^{-4}\,h^{3}\,{\rm Mpc}^{-3}$. The cuts have been applied so as to eliminate any overlaps between the two galaxy samples.}
    \label{fig:DESI_selection}
\end{figure*}

In Figure~\ref{fig:projected_DR7_colour_LGalaxies}, we found that galaxies in the mass range $9.0~\leq~\log\left[M_\star/{\rm M}_\odot\right]\leq10.5$ show the largest difference in clustering between MTNG740 and \lgal{} -- particularly for red galaxies. Focussing on this mass range in Figure~\ref{fig:colour_distr} (first three panels), we see interesting differences in the $g-r$ colours of objects in MTNG740 and \lgal{}. In particular, MTNG740 shows a prominent excess of red satellites (with respect to \lgal{}). This excess abundance of red satellites in this mass range is likely what contributes to the excess clustering at $r_p\lesssim3$~Mpc seen in Figure~\ref{fig:projected_DR7_colour_LGalaxies}. We also note some additional differences in the three most massive bins where, in particular, MTNG740 exhibits an extended blue tail, which is not seen in either \lgal{} or SDSS galaxies. This may be remedied, at least partially, by accounting for extinction by dust, which we have not done for MTNG740. The paucity of blue galaxies in the mass range $\log\left[M_\star/{\rm M}_\odot\right]\geq11.0$ results in the rather noisy clustering measurements in this mass range (bottom right panel in Figure~\ref{fig:projected_DR7_colour_LGalaxies}). Finally, although not shown here, we have checked that the red-blue galaxy cross-correlation in \lgal{} tracks the geometric mean of their individual auto-correlations to smaller scales than in MTNG740. 

\subsection{The clustering properties of DESI-like samples in MTNG740 and \lgal{}}
\label{sec:DESI}

An important application of the MTNG suite of simulations is to make predictions for ongoing and future galaxy redshift surveys; this is the focus of the present section. In particular, we consider populations of simulated galaxies that are representative of Luminous Red Galaxies (LRGs) and Emission Line Galaxies (ELGs) as will be targeted by the DESI survey. LRG-like galaxies identify massive, mostly quiescent galaxies hosted in group- and cluster-mass haloes ($M_{{\rm host}}\geq10^{13}\,{\rm M}_\odot$). The ELG sample targets highly star forming galaxies characterised by strong emission lines in their spectra which signifies the presence of young stars; these are typically found within less massive hosts than their LRG counterparts ($M_{{\rm host}}\sim10^{12}\,{\rm M}_\odot$). There has been significant recent interest in the relationship between ELG/LRG populations and their host haloes; for more details, we refer the reader to works by \cite{Geach2012,GonzalezPerez2018,Hadzhiyska2021,Hernandez-Aguayo:2020oiw,Jimenez2021,Yuan2022,Delgado2022a,Osato2022}.

In the present analysis, we focus mainly on ELG-like and LRG-like populations at $z=0.81$, comparable to the redshift at which the DESI ELG sample peaks in number density. We generate these samples with a target number density of $\overline{n}_{{\rm gal}} = 7\times10^{-4}\,h^{3}\,{\rm Mpc}^{-3}$ which, at $z=0.81$, is comparable to the target number density of ELGs in eBOSS \citep{Raichoor2017} and DESI \citep{DESI2016}, as follows:
\begin{itemize}
    \item {\bf ELGs}: we select galaxies with minimum stellar mass of $\left[1.06\times10^{10}\,{\rm M}_\odot, 1.2\times10^{10}\,{\rm M}_\odot\right]$ and minimum specific star formation rate (sSFR) $\left[4.7\times10^{-10}\,{\rm yr}^{-1}, 5.4\times10^{-10}\,{\rm yr}^{-1}\right]$ in MTNG740 and \lgal{}, respectively (in order to attain the same number density, slightly different thresholds are required for the two different galaxy formation models).
    \item {\bf LRGs}: we rank order galaxies in descending order of stellar mass and select the first N galaxies so as to reporduce the observed number density. This results in a minimum galaxy mass of $8.7\times10^{10}\,{\rm M}_\odot$ in MTNG740 and $7.4\times10^{10}\,{\rm M}_\odot$ in \lgal{}. We also impose a {\it maximum} sSFR cut equal to the minimum sSFR threshold for the ELG populations in order to guarantee that we include only massive quiescent galaxies \citep[see also][]{Boryana2022a,Boryana2022b}.
\end{itemize}
We note that the selection criteria described above identify ELG-like and LRG-like galaxies only approximately. In particular, we do not attempt to forward model photometry for MTNG galaxies into DESI photometric bands, where, for example, W1 luminosities and magnitudes are used to identify LRGs in the survey. We leave a full treatment of this kind to a future investigation. Instead, the cuts used in this paper are based on criteria identified by \cite{Hadzhiyska2021} after forward modelling galaxy SEDs in TNG-300-1.

Figure~\ref{fig:DESI_selection} shows the specific star formation rate as a function of stellar mass at $z=0.81$ in MTNG740 (left panel) and \lgal{} (right panel). The subset of galaxies identified as LRG-like and ELG-like (based on the criteria listed above) are highlighted using red and blue colours, respectively. By imposing a specific star formation rate cut for both selections we ensure that there is no overlap between the two populations of galaxies. The \lgal{} model, being semi-analytic in nature, is less restricted by numerical resolution and allows a denser sampling of galaxies less massive than $\sim10^{10}\,{\rm M}_\odot$. Note that because we have generated samples at the same number density within volumes of the same size ($740^3\,$~Mpc$^3$), we are left with the same total number of ELG-like and LRG-like galaxies in MTNG740 and \lgal{}.

\begin{figure}
    \centering
    \includegraphics[width=\columnwidth,trim={0.3in 0 0 0 },clip]{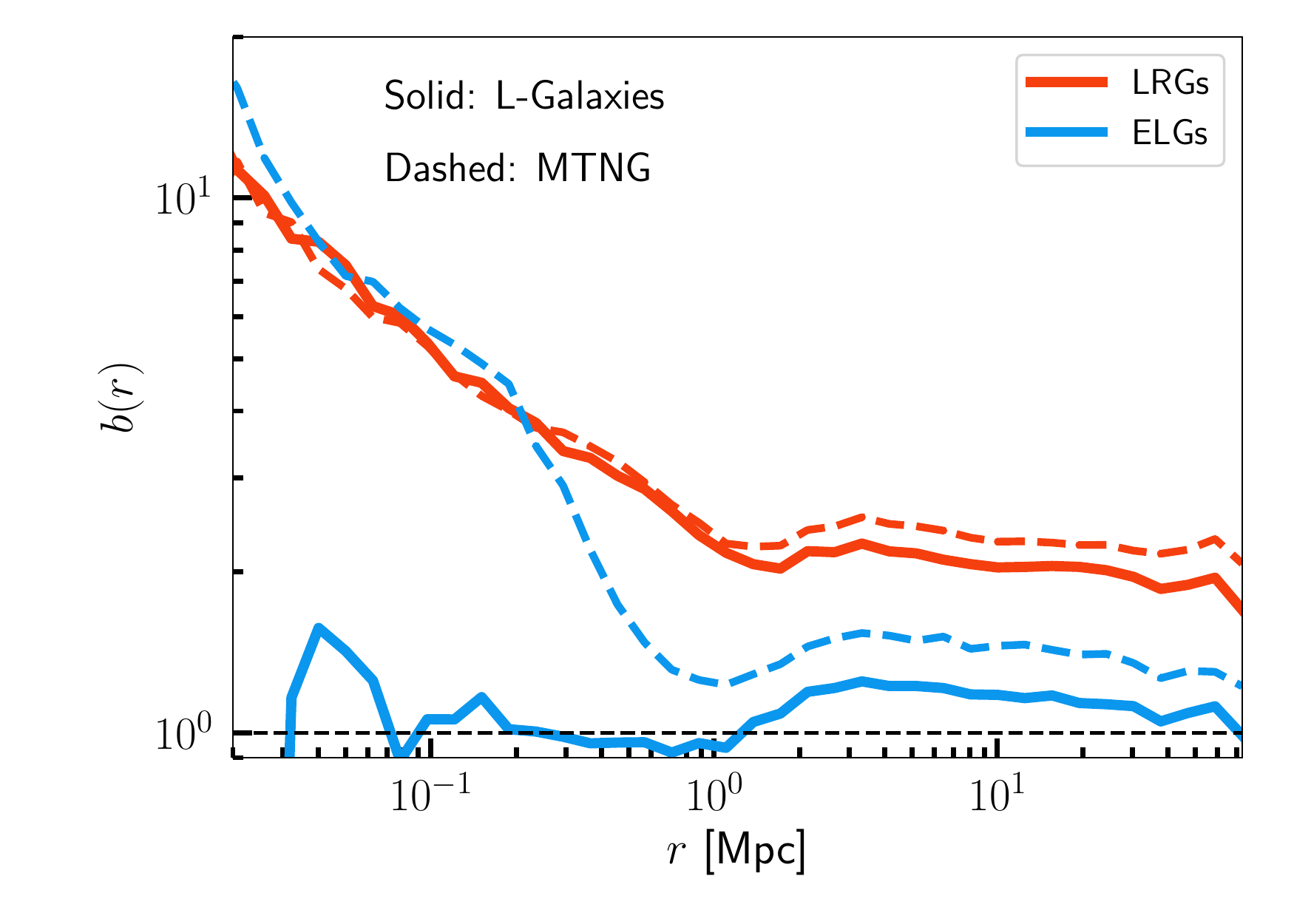}
    \caption{The scale-dependent galaxy bias, $b(r)$, as a function of 3D galaxy separation, $r$, for our DESI-like LRG (red) and ELG (blue) samples selected at $z=0.81$ and number density $\overline{n}_{{\rm gal}} = 7\times10^{-4}\,h^{3}\,{\rm Mpc}^{-3}$ (see also Figure~\ref{fig:DESI_selection}). The dashed curves are the results from MTNG740, while the solid curves are from \lgal{} run on the MTNG740-4320-A catalogue.  While there is good agreement between MTNG740 and \lgal{} across all scales for the LRG-like sample, a significant difference is observed in the one-halo regime for the bias of ELG-like galaxies. In particular, the strong scale dependence observed in MTNG740 is not reproduced in \lgal{}. We also observe a small offset in the two-halo regime. In all cases, the scale dependence of the galaxy bias settles down for separations larger than $10\,{\rm Mpc}$ or so, although it does not quite become constant.}
    \label{fig:bias_DESI}
\end{figure}

Having defined an appropriate sample of LRG-like and ELG-like galaxies, we are now in a position to predict the clustering properties of these populations. Rather than show the clustering strength, $\xi(r)$, it is interesting to consider the clustering {\it bias} of these objects. The scale-dependent bias of a tracer galaxy population, $g$, is defined as:
\begin{equation}
    b(r) = \sqrt{\frac{\xi_{gg}(r)}{\xi_{{\rm DM}}(r)}},
\end{equation}
where $\xi_{gg}(r)$ and $\xi_{{\rm DM}}(r)$, respectively, are the two-point correlation functions of the tracer population and the dark matter density field. 

Figure~\ref{fig:bias_DESI} shows the bias of our ELG- (blue) and LRG-like (red) samples in MTNG740 (dashed lines) and \lgal{} (solid lines). There are a number of interesting conclusions to be drawn from this diagram. First, it is clear to see a strong scale dependence exhibited by both galaxy types. The scale dependence is especially pronounced on small scales ($r<5$~Mpc); on larger scales, the bias becomes roughly constant. This is in line with the expectation that the large-scale galaxy bias is scale independent, making perturbative approaches to modelling galaxy bias appropriate on these scales. We also see that the large-scale bias for LRG-like galaxies is about a factor of two stronger than for ELGs; this is also expected as LRG-like galaxies are hosted in more massive haloes, which are themselves clustered more strongly than low mass objects. We find that the large-scale bias settles down at around $b(r)\approx2$ for LRGs and $b(r)\approx1.0$-$1.5$ for ELGs. Significant scale dependence in the galaxy bias begins to manifest in the transition regime between the one- and two-halo terms, where the clustering begins to be dominated by satellite galaxies. Trends similar to what we have shown here are also observed in TNG-300 \citep[see, e.g. Figure~19 in][]{Springel2018}.

\begin{figure}
    \centering
    \includegraphics[trim={4mm 0 0 0},clip,width=\columnwidth]{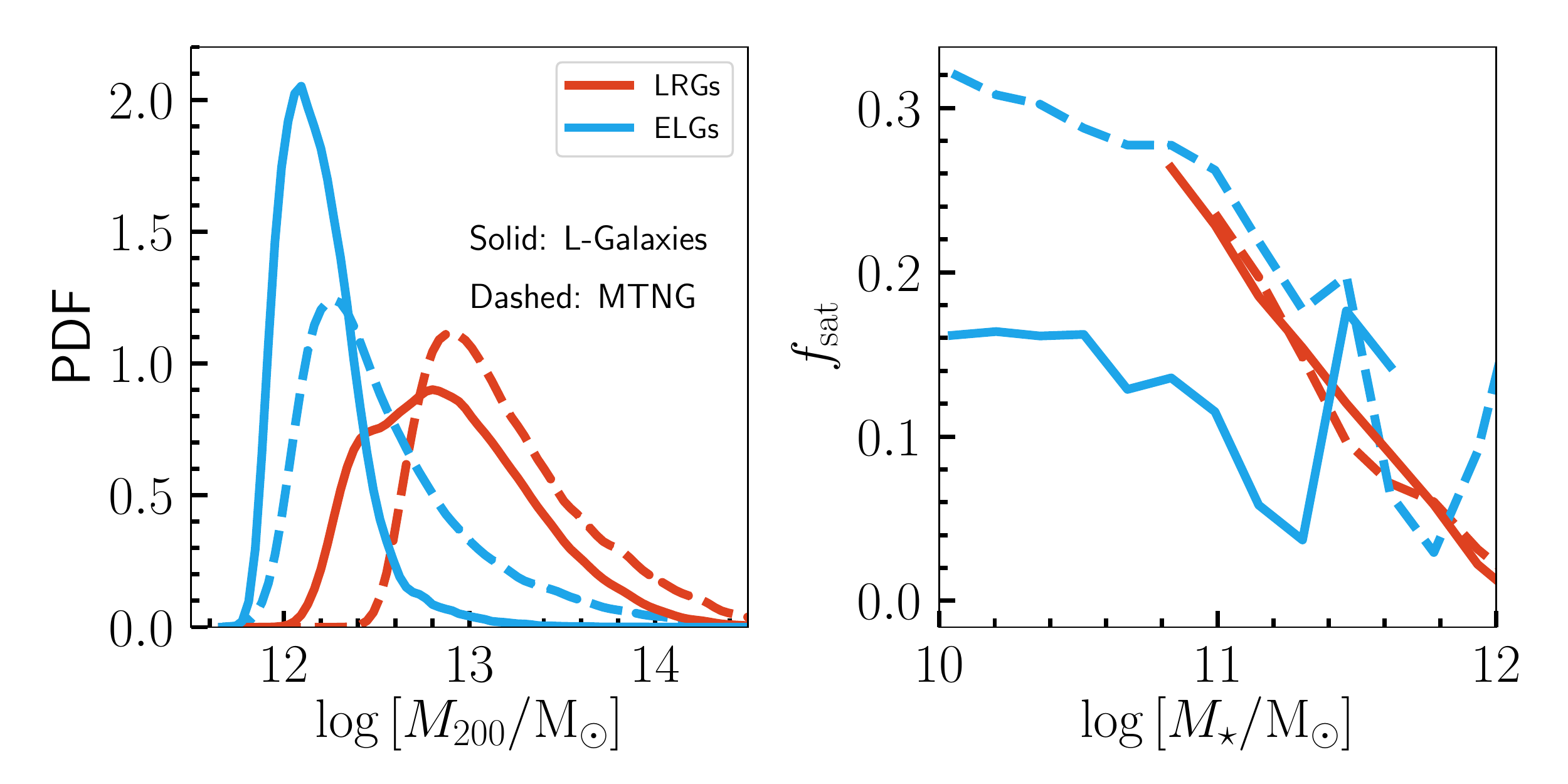}
    \caption{{\bf Left panel}: the distribution of halo masses hosting ELG and LRG-like galaxies at $z=0.81$ as identified in \lgal{} (solid lines) and MTNG740 (dashed lines). Qualitatively, both sets of galaxy catalogues predict that LRG-like galaxies are hosted in more massive haloes ($M_{200} \sim 10^{13}\,{\rm M}_\odot$) than ELG-like galaxies (typical host mass $M_{200}\sim 10^{12}\,{\rm M}_\odot$). Interestingly, while the distribution of halo masses for the LRG-like sample is broader in \lgal{}, it is the opposite for the ELG-like sample, where MTNG740 shows a wider distribution of host halo masses. {\bf Right panel}: The fraction of galaxies identified as satellites as a function of galaxy stellar mass, $M_\star$, at $z=0.81$. Colours and line styles are identical to those in the left panel. \lgal{} and MTNG740 predict almost identical satellite fractions for the LRG-like sample, but the fraction of ELG-like satellite galaxies in MTNG740 is typically double that of the \lgal{} catalogue. This is likely part of the reason that explains the difference in the small scale bias of the ELG-like population (see Figure~\ref{fig:bias_DESI}).}
    \label{fig:mhalo_distr}
\end{figure}

The comparison between MTNG740 and \lgal{} for the two galaxy types is also interesting. For the LRGs, we find excellent agreement between the two catalogues for galaxy separations of $100$~kpc all the way up to $75$~Mpc (1/10th of the box size, which is the maximum separation we consider when computing $\xi(r)$). For the ELG-like population, on the other hand, the behaviour is similar on scales larger than $r\sim2$~Mpc, but significantly different on smaller scales. While ELG-like galaxies in MTNG740 show strong scale-dependent bias (not dissimilar to the LRGs), this scale-dependence is largely absent in \lgal{}.

In order to explain the possible cause of these differences, it is instructive to examine the galaxies and host haloes of the LRG-like and ELG-like samples we have generated from each of MTNG740 and \lgal{}. The left panel in Figure~\ref{fig:mhalo_distr} shows the distribution of halo masses (in terms of $M_{200}$) of galaxies identified as LRG-like or ELG-like in MTNG740 (dashed lines) and \lgal{} (solid lines). Consistent with other works studying these populations, we find that the typical host halo mass for LRG-like galaxies is $M_{200}~\approx~10^{13}\,{\rm M}_\odot$ and $M_{200}~\approx~10^{12}\,{\rm M}_\odot$ for ELG-like galaxies. In fact, we largely do not find any LRGs hosted in the mass scale typical for ELG hosts. It is also interesting to note that in \lgal{}, ELG-like galaxies are hosted in a narrower mass range (centred around $10^{12}\,{\rm M}_\odot$) than in MTNG740. In the latter, for example, there is a non-negligible number of galaxies identified as ELGs hosted in haloes with $M_{200}\gtrsim10^{13}\,{\rm M}_\odot$ whereas we typically do not find ELG-like galaxies in hosts as massive as these in \lgal{}.

\begin{figure*}
    \centering
    \includegraphics[trim={2in 2in 0 0},clip,scale=0.4]{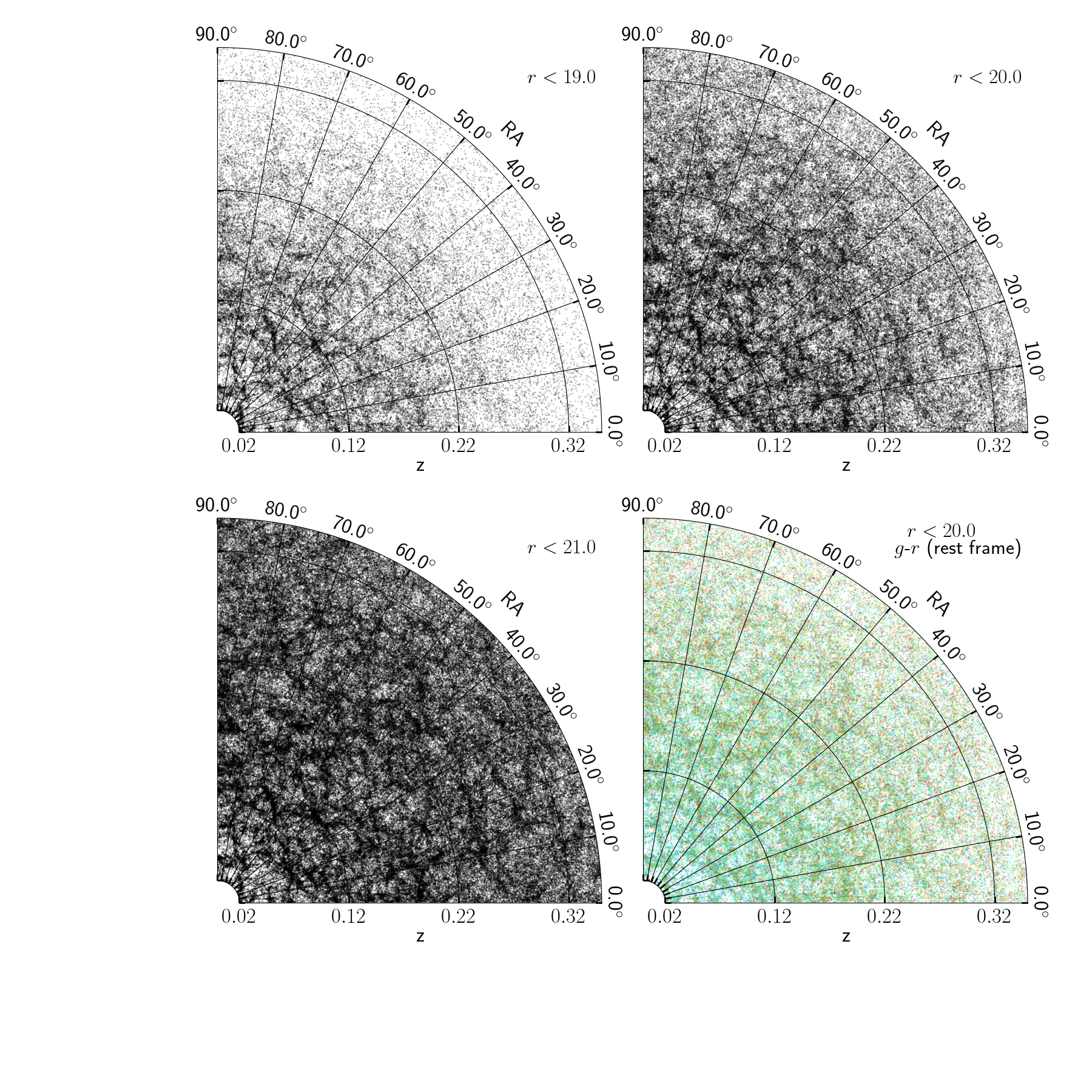}
    \caption{Slices through the MTNG740-2160-A \lgal{} lightcone catalogue. We display galaxy samples based on three separate $r$-band apparent magnitude cuts: $r<19$, $r<20$, and $r<21$ between $0.02<z<0.35$. Each subsequent cut includes fainter galaxies as revealed by the increased density of structure from the top-left to the bottom-left panels. The last panel (bottom right) again shows the $r<20$ sample, with individual galaxies now coloured by their $g$-$r$ colour. To avoid overcrowding, we only show galaxies between $0^\circ < {\rm DEC} < 20^\circ$ and $0^\circ < {\rm RA} < 90^\circ$.}
    \label{fig:lightcones}
\end{figure*}

To understand this further, in the right-hand panel in Figure~\ref{fig:mhalo_distr} we show the fraction of galaxies of a given stellar mass, $M_\star$, that are identified as {\it satellite} galaxies ($f_{{\rm sat}}$). We remind the reader that a galaxy is identified as a satellite when it is not formed inside the most massive subhalo within a given host halo. Focussing on the LRG-like populations to begin with, we find  near-identical behaviour in MTNG740 and \lgal{}. As expected, galaxies of increasing stellar mass are less likely to be identified as satellites; the most massive galaxies are more likely to be formed within the principal subhalo within a host halo. On the other hand, around a quarter of galaxies with stellar mass $M_\star\sim10^{10.8}\,{\rm M}_\odot$ are satellite LRGs.

We find more significant differences when comparing the satellite fraction of the ELG-like populations. Here, we also note the trend of decreasing $f_{{\rm sat}}$ with increasing $M_\star$, though the dependence becomes gentler below $M_\star~\sim~10^{11}\,{\rm M}_\odot$. While a handful of galaxies with $M_\star>10^{11.6}\,{\rm M}_\odot$ are identified as satellite ELGs in MTNG740, there are no such objects in \lgal{}. This is consistent with our observation in the left-hand panel of the same figure, where it was found that \lgal{} has an absence of ELGs hosted in the most massive haloes. Shifting to lower mass galaxies, we find that in MTNG740, satellite ELGs appear twice as often at $M_\star~\sim~10^{10}\,{\rm M}_\odot$, and nearly four times as often at $M_\star~\sim~10^{11}\,{\rm M}_\odot$ than they do in \lgal{}. The considerable difference in satellite fractions provides a tantalising metric to potentially constrain models of galaxy formation with upcoming data. However, there is significant systematic uncertainty associated with inferring $f_{\mathrm{sat}}$ observationally as it depends sensitively on the details of the galaxy occupation model used to fit the clustering.

The comparatively low density of satellite ELGs in \lgal{} likely contributes to part of the reason why the small-scale bias, $b(r)$, is lower than in MTNG740 (see Figure~\ref{fig:bias_DESI}). The clustering of satellite galaxies dominates the clustering signal below $r~\lesssim~1~-~3$~Mpc, and it is on these scales where we notice the large divergence in the predicted bias measurement of the two sets of simulated ELG samples. This difference is reflective of the different HODs for emission line galaxies in \lgal{} compared to MTNG740. We tend to find more star-forming (bluer) central galaxies in the \lgal{} catalogue (see Figure~\ref{fig:colour_distr}), which means for a number density-limited sample the overall {\it fraction} of satellite ELGs is comparatively lower in \lgal{} than in MTNG740.

We now conclude our examination of the clustering properties of galaxies in the periodic MTNG740 and MTNG740-4320 \lgal{} volumes. A realistic galaxy redshift survey observes galaxies within a lightcone, with footprints and selection functions that are specific to individual surveys. In the following subsection, we will consider the clustering properties of galaxies identified on lightcones that have been constructed from our simulations on-the-fly.

\subsection{The clustering properties of mock catalogues on the lightcone}
\label{sec:lightcone}

In this subsection, we present clustering predictions of mock galaxy catalogues constructed on lightcones output as part of the MTNG suite. In what follows, we will consider \lgal{} catalogues constructed from the MTNG740-2160-A/B dark matter-only runs. The full details of the lightcone construction strategy can be found in the companion paper by \cite{Barrera2022}; here, we consider a particle lightcone occupying one octant of the sky in the redshift range $0<z<1.5$, resulting in a maximum comoving distance of 4.5~Gpc.

To demonstrate a potential application of our lightcones, we construct an SDSS-like magnitude limited sample of galaxies between $0.0<z<0.5$. The \lgal{} ligtcone outputs provide observer frame apparent $r$-band magnitudes that we use to generate such samples. Figure~\ref{fig:lightcones} shows slices through the resulting mocks generated from MTNG740-2160-A using three apparent magnitude cuts: $r<19$, $r<20$, and $r<21$. Each successive cut results in the inclusion of fainter galaxies, which is why the space density of galaxies increases from one wedge to the next. We display the positions of galaxies only within angular positions $0^\circ < {\rm DEC} < 20^\circ$ and $0^\circ < {\rm RA} < 90^\circ$ in order to avoid oversaturating the diagram. The structure of the cosmic web, traced by our semi-analytic galaxies, can be seen clearly, with filaments and walls reminiscent of SDSS structure.  The last panel in the bottom right of this figure shows the $r<20$ sample with galaxies coloured according to their $g$-$r$ colour. We see that the magnitude-limited sample picks up nearby faint, blue galaxies, and distant bright, red galaxies.

\begin{figure}
    \centering
    \includegraphics[width=\columnwidth,trim={0.3in 0 0 0 },clip]{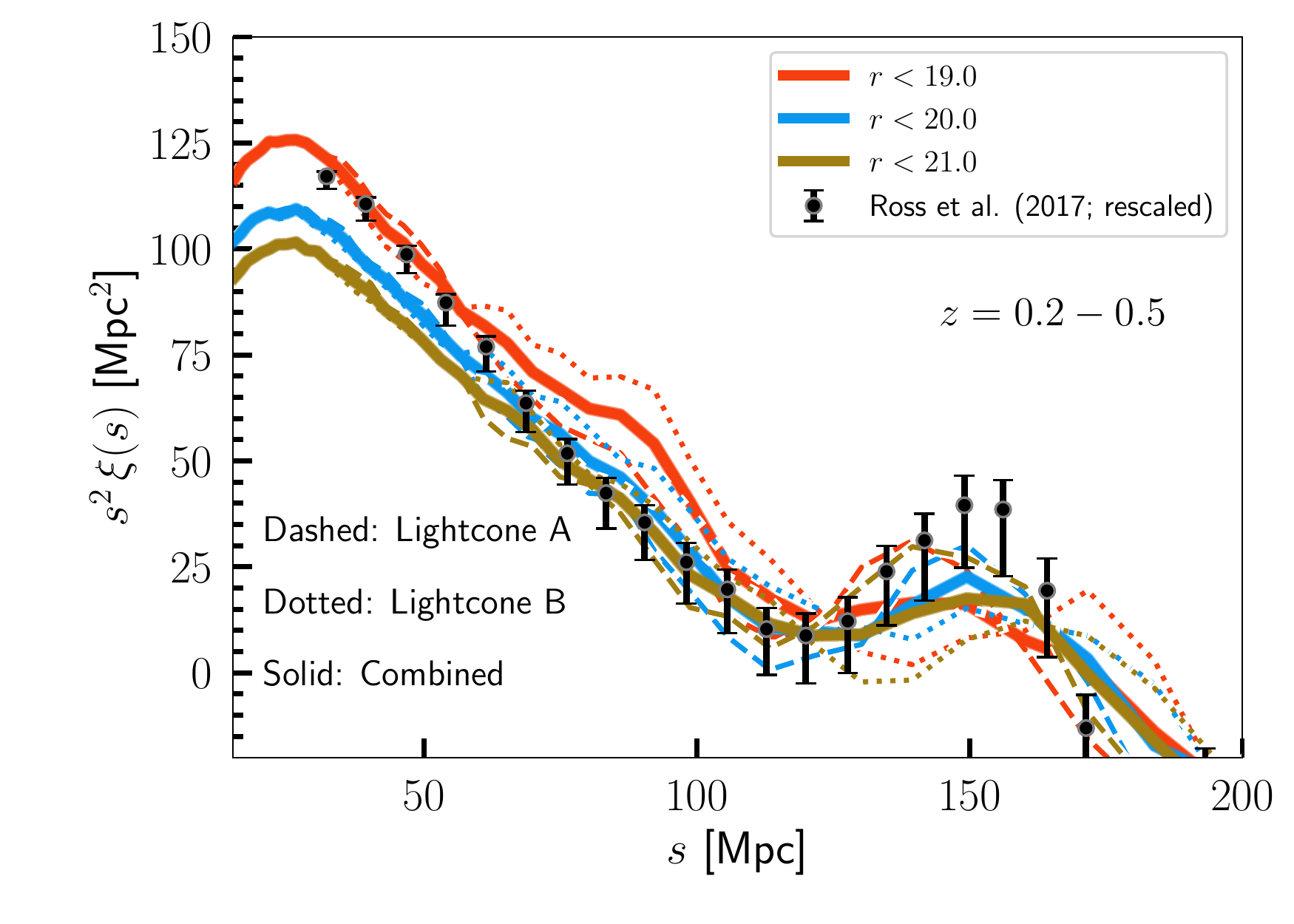}
    \caption{The redshift-space correlation function, $\xi(s)$, of galaxies identified between $0.2<z<0.5$ in the \lgal{} lightcone catalogues. Each line shows the auto-correlation function of galaxies selected by apparent $r$-band magnitude; the dashed and dotted curves, respectively, show the correlation functions measured from the A and B realisations of the MTNG740-2160 lightcones (see main text for details), while the thick solid line shows the {\it average} of the two. In order to emphasise features in $\xi(s)$, the correlation function has been multiplied by $s^2$, where is the galaxy separation in redshift-space. The black points with error bars represent the clustering measured in BOSS by \citet{Ross2017}. Note that the data have been shifted down by a constant factor to aid comparison with the simulation measurements.}
    \label{fig:xi_s}
\end{figure}

We are now in a position to compute the redshift-space clustering, $\xi(s)$, of galaxies defined on the lightcone (where $s$ is the galaxy pair separation calculated in redshift-space). We use the \cite{Landy1993} estimator, defined so that:
\begin{equation}
    \xi(s) = \frac{DD(s)-2DR(s)+RR(s)}{RR(s)},
\end{equation}
where $DD(s)$, $RR(s)$, $DR(s)$, respectively, are the data-data pair counts, random-random pair counts, and data-random cross pair counts computed in redshift space.

Figure~\ref{fig:xi_s} shows the redshift-space correlation function, $\xi(s)$, of galaxies in the redshift range $0.2<z<0.5$ in the mock catalogue defined above. Each curve shows the clustering of samples generated using different apparent $r$-band magnitude thresholds; the `A' and `B' realisations of the MTNG740-2160 \lgal{} lightcones are shown, respectively using the dashed and dotted lines. The correlation function has been multiplied by $s^2$ to exaggerate fluctuations in $\xi(s)$. We see the clear presence of a BAO feature in each of clustering samples occurring at scales of $s\approx150$~Mpc, as expected. We also find that the amplitude of the clustering varies between the samples: the $r<21$ includes many faint galaxies, whereas the $r<19$ sample includes only the brighter (and more biased) galaxies. The variance just between the `A' and `B' realisations is interesting to note, particularly in the exact location of the BAO feature \citep[see also][for a more detailed analysis of this effect]{Barrera2022}. The thick solid line combines the two realisations to yield a smoother clustering measurement with suppressed noise on large scales, showing the advantage of combining volumes with `fixed' and `paired' initial conditions. 

For comparison, we have also included the clustering measured in the Baryon Acoustic Oscillation Spectroscopic Survey \citep[BOSS,][]{Reid2016} by \cite{Ross2017} for galaxies in the redshift range $0.2<z<0.5$; this is shown using the black points with error bars. Note that the \cite{Ross2017} measurements have been shifted down by a constant factor so as to enable comparison with the simulations. We find  reasonable qualitative agreement between the clustering measured in BOSS and our magnitude-limited samples, although there are some differences to note, particularly in the amplitude of the BAO peak. Some differences are inevitable, since we have taken no great care to match either the colour selection (where, in addition to a brightness cut, colour cuts in the $g$, $r$, and $i$-bands are also used), nor do we attempt to mimic the redshift-dependent galaxy number density in BOSS.

The peculiar motions of galaxies in redshift-space add a contribution to their observed redshift (i.e., their position along the radial direction) that can be attributed to the growth rate of structure on large scales. These redshift space distortions (RSDs) propagate through to the clustering measured from galaxies in a survey, making it more anisotropic. By measuring the degree of anisotropy, therefore, it is possible to quantify the characteristic amplitude of pairwise peculiar motions of galaxies and, therefore, the growth rate of structure \citep[e.g.][]{Kaiser1987,Hawkins2003,Zehavi2005,Samushia2012,Reid2012}. A convenient way to express the anisotropy in redshift-space clustering is to decompose $\xi(s)$ into its Legendre multipoles \citep{Hamilton1992}:
\begin{equation}
    \xi_\ell(s) = \left(2\ell+1\right) \int_0^1 \xi(s,\mu) P_\ell(\mu)\,{\rm d}\mu,
\end{equation}
in which $s=\left|{\bf s}\right|$ is magnitude of the pair separation vector, ${\bf s}$, $\mu=\upi/s$ is the cosine of the angle between the line-of-sight direction and ${\bf s}$, and $P_\ell(\mu)$ is the $\ell^{{\rm th}}$ order Legendre polynomial. Due to symmetry, the first three non-zero Legendre multipoles are those of order $\ell~=~0,2,4$ which, respectively, are referred to as the monopole ($\xi_0(s)$), quadrupole ($\xi_2(s)$), and hexadecapole ($\xi_4(s)$) terms of the redshift-space correlation function. In the absence of RSDs, the quadrupole and hexadecapole terms vanish.

\begin{figure}
    \centering
    \includegraphics[width=\columnwidth]{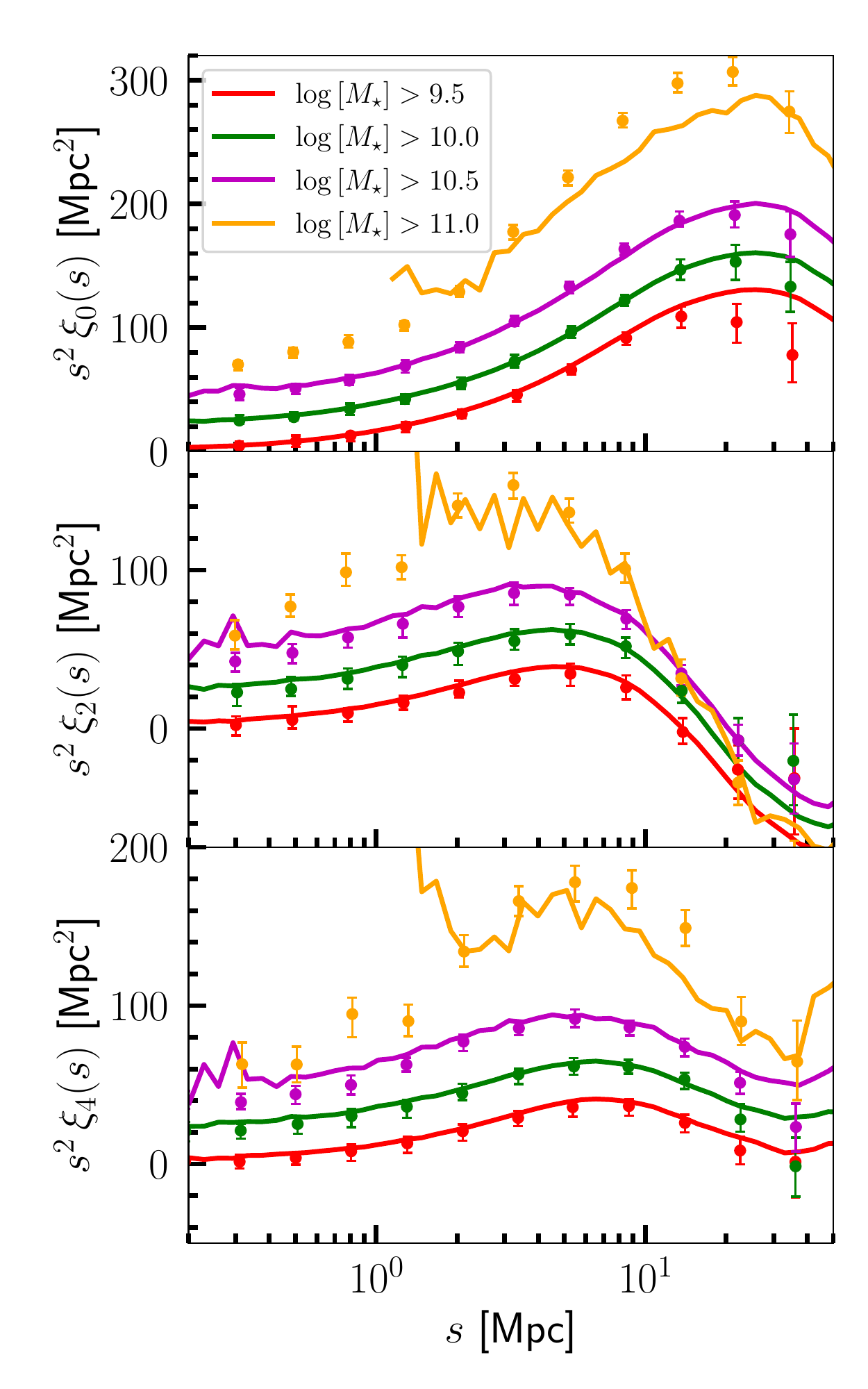}
    \caption{The monopole ($\xi_0(s)$, top panel), quadrupole ($\xi_2(s)$, middle panel), and hexadecapole ($\xi_4(s)$, bottom panel) of the redshift-space correlation function measured from the \lgal{} lightcone catalogues. Each curve shows the result after combining the A and B realisations of the MTNG740-2160 lightcones. The different colours represent the multipoles of the correlation function for galaxies selected by according to stellar mass, $M_\star$. The symbols with error bars show the clustering measured in SDSS by \citet{DongPaez2022}; we have followed the convention of these authors and offset each set of curves vertically in steps of $20\,{\rm Mpc}^2$ for clarity.}
    \label{fig:xi_multipole}
\end{figure}

In Figure~\ref{fig:xi_multipole}, we present predictions for the monopole, quadrupole, and hexadecapole of the redshift-space correlation function measured from our lightcone mock catalogues. Once again, we have combined the individual measurements from the `A' and `B' realisations. We show the measurements for samples now constructed at fixed {\it stellar mass} thresholds: $\log M_\star > 9.5$ (red); $\log M_\star > 10.0$ (green); $\log M_\star > 10.5$ (purple); $\log M_\star > 11.0$ (orange). Each curve has been offset vertically in steps of $20\,h^{-2}\,{\rm Mpc}^2$ (starting from the lowest mass bin) to aid clarity. We also show (using points with errorbars in corresponding colours) the redshift-space multipoles measured in SDSS data by \cite{DongPaez2022}; these data have also been offset like the curves.

As expected, we find non-zero contributions to $\xi_2(s)$ and $\xi_4(s)$ through the inclusion of galaxy peculiar motions in the RSD. The strength of the signal is largest towards small galaxy separations, where peculiar motions are enhanced due to non-linear growth. Note that the RSDs also have a non-negligible impact on the monopole term. On large scales, the quadrupole term is expected to become negative\footnote{This effect is not so obvious in Figure~\ref{fig:xi_multipole} due to the vertical offsets we have applied between the curves.} as a reflection of the infalling motions of galaxies on the outskirts of clusters; this causes a compression of structure along the line-of-sight. On smaller scales, the quadrupole term is again positive due to the virialised motions of galaxies within clusters causing a net elongation of structures along the line-of-sight (the so-called ``fingers of God'' effect). The hexadecapole term, $\xi_4(s)$, encodes similar information, but now regarding paired infalling galaxies.

Figure~\ref{fig:xi_multipole} shows excellent agreement between the SDSS redshift-space multipoles and the predictions of our lightcone catalogues, all the way from separations of $s\sim200$~kpc to $~50$~Mpc. The measurements for the largest mass bin are noisy, and the agreement here is not as good for lower mass bins; it is likely that our (simulated) sample size is statistically limited in this regime. It will be interesting to revisit this further using lightcones generated from the 3~Gpc MillenniumTNG-XXL volume, which boasts more than 60 times the volume of the runs presented in this paper. We set aside this investigation as future work. 

\section{Discussion \& Conclusions}
\label{sec:conclusions}

This work is part of a series of introductory papers comprising the MillenniumTNG (MTNG) project, a new simulation series that blends the successes of the IllustrisTNG (TNG) galaxy formation model \citep{Weinberger2017,Pillepich2018} with high-resolution, large volume simulations. Our main objective in this work was to present the clustering predictions of galaxies extracted from the flagship full physics simulation, MTNG740, which evolves a 740~Mpc periodic volume with a baryonic mass resolution of $3\times10^7\,{\rm M}_\odot$. This mass resolution has been shown to be sufficient to achieve converged properties in the regime relevant to large-scale structure analyses \citep{Pakmor2022}, which is the primary domain of applications anticipated for MTNG. The clustering predictions for these galaxies are considered alongside a synthetic catalogue generated using the \lgal{} semi-analytic model of galaxy formation, run on merger trees extracted from the dark matter-only counterpart to MTNG740. Our main findings are summarised below:
\begin{enumerate}
    \item We find that at $z=0.1$, the projected two-point clustering of galaxies, $w_p(r_p)$, extracted from MTNG740 is in good agreement with that of observed galaxies (Figure~\ref{fig:projected_DR7}). In particular, measurements of $w_p(r_p)$ as a function of galaxy stellar mass show a reasonable match to the measurements from SDSS data by \cite{Guo2011}. The clustering is also shown to be well-converged against the higher resolution TNG-300-1 simulation, in which the baryonic mass resolution is roughly twice that in MTNG740.
    \item When split by $g-r$ colour, the projected clustering shows good agreement with the observed data for blue galaxies, but MTNG740 overpredicts the clustering observed for red galaxies, particularly for objects in the mass range $9.5\leq\log\left[M_\star/{\rm M}_\odot\right]\leq10.5$ and in the one-halo regime (Figure~\ref{fig:projected_DR7_colour}). This suggests an overabundance of quenched, red satellite galaxies in our full physics model. This problem is almost absent in the semi-analytic model.
    \item A consequence of this is that the predicted {\it cross-correlation} of red and blue galaxies in MTNG740 shows a  suppression relative to the geometric mean of their individual auto-correlation functions (Figure~\ref{fig:colour_crosscorrelation}). In galaxies less massive than $\log\left[M_\star/{\rm M}_\odot\right]\lesssim10.5$, the cross-correlation effectively flattens out below $r_p\sim3$~Mpc. This potentially implies a significant separation of red- and blue-satellite hosting haloes in this mass range. Again, this issue is absent in the semi-analytic model.
    \item The much better agreement of \lgal{} model with observed colour-dependent clustering as a function of stellar mass may in part be because the observed active/passive fractions of galaxies (though not their clustering) were used to calibrate the semianalytic  model.  No such explicit requirement was imposed on the MTNG full physics run (or rather on the ealier TNG runs from which the MTNG parameters were taken). Indeed, comparing the galaxy colour distributions in the simulated catalogues, we find a clear excess peak of red satellites in MTNG740 (Figure~\ref{fig:colour_distr}).
    \item We then consider the clustering predictions of galaxy samples from each of MTNG740 and \lgal{} chosen so as to reflect the selection of luminous red galaxy (LRG) and emission line galaxy (ELG) samples as targeted by the DESI survey (Figure~\ref{fig:DESI_selection}). Our models predict that the large-scale bias, $b(r)$, of these samples becomes roughly linear and scale-independent on scales larger than $r\sim10$~Mpc, with the value settling around $b(r)\approx2$ for the LRGs and $b(r)\approx1$-$1.5$ for ELGs (Figure~\ref{fig:bias_DESI}). 
    \item In \lgal{ } the small-scale bias of ELG-like samples is much weaker than in MTNG740; this is due to the fact that a much smaller fraction of low-mass ELG-like galaxies are satellites in the semi-analytic model, around $2-4$ times fewer than in the MTNG740 full physics run (Figure~\ref{fig:mhalo_distr}). Both models predict that LRG-like galaxies are hosted in haloes with mass $\sim10^{13}\,{\rm M}_\odot$ while ELGs tend to live in haloes that are an order of magnitude less massive. The distribution of halo masses hosting ELG-like galaxies is narrower in \lgal{} than in MTNG740. On the other hand, the bias for LRG-like galaxies is similar across all scales in both models. 
    \item We created mock galaxy samples from the lightcone output generated on-the-fly in MTNG. We generated mocks selected above three apparent $r$-band magnitude thresholds (Figure~\ref{fig:lightcones}), and presented the redshift-space clustering predictions, $\xi(s)$, for galaxies in this lightcone between $0.2<z<0.5$ (Figure~\ref{fig:xi_s}). We are able to clearly identify the BAO feature in each of these samples.
    \item Finally, we concluded our investigation by decomposing the redshift-space clustering into Legendre multipoles, studying the monopole, quadrupole, and hexadecapole terms (Figure~\ref{fig:xi_multipole}). Upon comparing the predictions of our mock lightcone catalogues with stellar mass-selected samples in SDSS, we find good agreement with observations, thus signifying that our lightcone mocks contain reasonably realistic redshift space distortions.
\end{enumerate}
Our results show the important insights into galaxy formation that may be inferred from studies of galaxy clustering. By considering the clustering predictions of galaxies selected according to various types, and measured across a wide range of spatial scales, we are able to gain a more comprehensive understanding of how details of the galaxy formation process manifest in the evolved galaxy density field we observe today. With ongoing surveys like DESI, as well as upcoming ones like {\it Euclid} and the {\it Roman Space Telescope}, the opportunity to use high-precision measurements of galaxy clustering to constrain how galaxies form will only improve. This makes it an exciting time to also push cosmological hydrodynamical and semi-analytic simulations to the regime relevant to these surveys. The results presented in this paper are just a small representation of the kinds of applications the MillenniumTNG project is primed for. Further demonstrations of the richness of this data set will be the subject of future work. 

\section*{Acknowledgements}

We are thankful to the referee for their careful reading of this manuscript and for making suggestions that improve the quality of our work. The authors gratefully acknowledge the Gauss Centre for Supercomputing (GCS) for providing computing time on the GCS Supercomputer SuperMUC-NG at the Leibniz Supercomputing Centre (LRZ) in Garching, Germany, under project pn34mo. This work used the DiRAC@Durham facility managed by the Institute for Computational Cosmology on behalf of the STFC DiRAC HPC Facility, with equipment funded by BEIS capital funding via STFC capital grants ST/K00042X/1, ST/P002293/1, ST/R002371/1 and ST/S002502/1, Durham University and STFC operations grant ST/R000832/1. SB is supported by the UK Research and Innovation (UKRI) Future Leaders Fellowship [grant number MR/V023381/1]. CH-A acknowledges support from the Excellence Cluster ORIGINS which is funded by the Deutsche Forschungsgemeinschaft (DFG, German Research Foundation) under Germany’s Excellence Strategy – EXC-2094 – 390783311. VS and LH acknowledge support by the Simons Collaboration on “Learning the Universe”. LH is supported by NSF grant AST-1815978. 

\section*{Data Availability}

The MillenniumTNG simulations will be made fully publicly available at \url{https://www.mtng-project.org/} in 2024. The data shown in the figures of this article will be shared upon reasonable request to the corresponding author.



\bibliographystyle{mnras}
\bibliography{mtng_clustering} 



\bsp	
\label{lastpage}
\end{document}